\documentclass[pra,final,twocolumn,showpacs,aps]{revtex4-1}
\usepackage{multirow}
\usepackage{graphicx}
\usepackage{amssymb}
\usepackage{amsmath}
\usepackage{color}
\usepackage{xcolor}
\usepackage{multirow}
\usepackage{afterpage}
\usepackage{hyperref}
\usepackage{float}
\usepackage{ulem}

\begin{document}
\title[Qt in DBEC]{
Vortex dynamics and turbulence in dipolar Bose-Einstein condensates}
\author{S. Sabari$^{(a)}$\email{ssabari01@gmail.com}, 
R. Kishor Kumar$^{(a)}$\email{kishor.bec@gmail.com}
and Lauro Tomio$^{(a,b)}$\email{lauro.tomio@unesp.br}}
\affiliation{{$^{(a)}$Instituto de F\'\i sica Te\'{o}rica, Universidade 
Estadual Paulista, 01140-070 S\~{a}o Paulo, SP, Brazil\\}
{$^{(b)}$Centro Internacional de Física, Instituto de Física, Universidade de Brasília, 70910-900 Brasília, DF, Brazil}
}\date{\today}
\begin{abstract}
Quantum turbulence indicators in dipolar Bose-Einstein condensed fluids, following emissions of 
vortex-antivortex pairs generated by a circularly moving detuned laser, are being provided 
by numerical simulations of the corresponding quasi-two-dimensional Gross-Pitaevskii formalism with repulsive 
contact interactions combined with tunable dipole-dipole strength.
The critical velocities of two variants of a circularly moving obstacle are determined and analyzed for 
vortex-antivortex nucleation in the form of regular 
and cluster emissions. The turbulent dynamical behavior is predicted to follow closely 
the initial emission of vortex-antivortex pairs, relying on the expected 
Kolmogorov's classical scaling law, which is verified by the spectral analysis of 
the incompressible part of the kinetic energy.  
Within our aim to provide further support in the up-to-now investigations of quantum turbulence, 
which have been focused on non-dipolar Bose-Einstein condensates,
we emphasize the role of dipole-dipole interactions in the fluid dynamics.
\end{abstract}
\maketitle
\section{Introduction}
\label{sec1}
Dipole-dipole interactions, which are manifested between particles with permanent electric 
or magnetic dipoles, have attracted a lot of interest in cold-atom physics. They are expected to 
lead to novel kinds of degenerate quantum gases even in the weakly interacting limit. The
theoretical foundation with related progress can be found in review articles, as in  
Refs.~\cite{Baranov2008,Lahaye2009}.
The control of effective atom-atom dipole interactions under reasonable laboratory conditions
was shown to be possible in Ref.~\cite{1998Marinescu} and following seminal theoretical works 
reporting the tunability of such interactions~\cite{2001Yi,2002Santos,Giovanazzi2002}. 
Soon after that, it was reported the remarkable observations of dipolar Bose-Einstein 
condensates (BECs) with isotopes of chromium ($^{52}$Cr)~\cite{Griesmaier2006,Lahaye2007,2008Koch}, 
dysprosium ($^{164}$Dy)~\cite{Lu2011,Youn2010} and erbium 
($^{168}$Er)~\cite{Aikawa2012,Chomaz2016}. 
Along with this new exciting branch for investigations opened in cold-atom
physics~\cite{Lahaye2009,Baranov2012,Fetter2009,2022Klaus,Chomaz2023,2023Bland},
it was recently reported the production of another dipolar BEC with an isotope of 
europium ($^{151}$Eu)~\cite{2022Miyazawa}.
{ Besides their quite different characteristics, the dipole-dipole and contact $s-$wave 
interactions provide the leading order nonlinear effects, with the dipole-dipole 
interactions (DDI) being anisotropic with long-range behavior, opposing the short-range 
contact interactions. In the case of polar atoms, both can be varied widely, from 
repulsive to attractive, such that they are convenient parameters to control  
experimental realizations of confined BECs. In order to manipulate $s-$wave contact 
interactions in cold-atom physics, Feshbach resonance mechanisms have been extensively 
employed~\cite{Feshbach1962}, since the first experimental observation of
these resonances in a BEC~\cite{1998Inouye}. When considering polar atoms, we can further 
control their interactions, either via the magnitude of the external (magnetic or electric)
field being applied, or by modulating the alignment of this field with the 
intrinsic atomic dipole moments, which allows tuning the magnitude and sign of the 
DDI~\cite{Giovanazzi2002,Kumar2019}. 
Concerned the stability of dipolar BECs, 
among the several studies following Ref.~\cite{2008Koch}, 
the dynamical stabilization was explored by time modulation of 
scattering length and using the interplay between nonlinear 
interactions ~\cite{Sabari2015,Sabari2018a,Sabari2022}.
Noticeable is the crescent interest in binary dipolar mixtures, 
exploring miscible and immiscible regimes~\cite{2012Wilson,2017KumarJPC}, with an experimental realization 
(using erbium and dysprosium) reported in \cite{2018Trautman}. Along with these studies, 
the intensive investigations on rotational properties and vortex dynamics with binary dipolar 
atoms~\cite{Yi2006,Wilson2010,Mulkerin2013,Zhang2016,Kumar2017,Sabari2017,Martin2017}
aim to provide a platform to access plenty of other many-body quantum phenomena, such as 
the possible creation of long-lived quantum-droplet states and expected connections 
with superfluidity in BEC.

Quantum vortices in a BEC~\cite{Fetter2009} can be created only at a minimum critical 
angular velocity, when energetically favorable, different from their occurrence 
in normal fluids. Within a realistic experiment, the critical rotation for vortex 
nucleation can be larger due to dynamical instabilities at the boundaries.
In superfluid and superconductor phase transitions, quantum vortices play
a prominent role, as known since discoveries related to Helium 
superfluidity~\cite{Schwarz1987} and at high-temperature 
superconductors~\cite{Blatter1994}. Beyond that, they are commonly 
observed and investigated in a wide range of contexts going beyond cold-atom 
physics, such as on exciton-polariton condensates~\cite{Lagoudakis2008}, on 
polariton superfluids~\cite{Sanvitto2010}, and in optics~\cite{Willner2012,Molina2007}.
Quantized vortices have been nucleated in BEC through several techniques, 
such as rotating the trapping potential or thermal cloud~\cite{2001Madison}
(in experiments consistent with following up numerical analysis~\cite{Tsubota2002,Sabari2019}),
stirring the condensate with a blue- or red-detuned laser 
beam~\cite{Raman1999,Raman2001,Neely2010,White2012,Stagg2015}, 
moving a condensate past a defect~\cite{Henn2009}, 
rapid quench phase transitions of a cooling condensate~\cite{Lamporesi2013}, 
or decay instability of a soliton~\cite{Navon2015}. More recently,
vortex nucleations have been applied to lattice configurations
\cite{2023Jezek}, by predicting vortex positions along low-density paths separating 
the sites.

Within a recently reported experiment, by applying a moving Gaussian obstacle in a 
BEC~\cite{Lim2022}, the authors have shown that a tangle of quantized vortices can 
be produced. As known from classical fluid dynamics, vortex tangles are understood 
as a signature of turbulence, with the structure of turbulence being already 
associated with the flow of incompressible viscous fluids by
Kolmogorov~\cite{Kolmogorov}. Since Onsager’s ground-breaking theoretical work 
linking turbulence with a point vortex dynamics in a two-dimensional (2D) 
fluid~\cite{Onsager1949}, it has been hoped that the simple fundamental rules behind 
quantum theories, together with the recent experimental advances in studying vortex
dynamics in superfluids, will aid in understanding the nature of turbulence. 
As pointed out in 1963 Feynman Lectures on Physics~\cite{Feynman}, the analysis of 
circulating turbulent fluids is one of the most important problems in nature, left over 
a hundred years, with nobody really been able to analyze it mathematically satisfactorily 
still to be solved in spite of its importance.
Subsequently, with expectation that some light on the general solution of classical 
turbulence can be found in the so-called {\it quantum turbulence} (QT), 
the similarities between classical turbulence with superfluidity and QT
have been explored in several works and reviews~\cite{Barenghi2001,2002Vinen,2013Skrbek,2014Mantia},
which are mainly concerned on the classical length scales large, as compared with the characteristic 
quantum length scale found by the spacing between vortex lines.

After many years of research with superfluid helium systems, QT became a 
well-established field for investigation, motivating numerous new insights and 
developments regarding its possible universality~\cite{Barenghi2001}. The discovery
of links between classical and quantum turbulence has remained a strong motivating 
factor for QT research, with particular interest in cold-atom physics, considering
the actual available experimental possibilities, being a platform for probing and 
studying superfluid flows~\cite{Kobayashi2007,Pethick2008}. 
In this regard, on the way to understanding the superfluidity phenomenon, 
the critical speed below which there is no longer dissipation in a fluid was studied
considering a  BEC experiment of sodium atoms~\cite{Raman1999}. The formation of 
clusters of like-sign vortices has been studied extensively in 2D quantum 
fluids~\cite{White2012,Reeves2014,Yu2016,Groszek2016,Salman2016,Gauthier2019,
Johnstone2019}, with the phenomenon being related to the inverse energy
cascade~\cite{Reeves2013,Kraichnan1967} and large structure produced as the 
energy is transferred from a small to a large spatial scale. The dynamical 
production and decay of turbulence and vorticity was studied in 
Ref.~\cite{Parker2005} by assuming a stirred atomic BEC, within an investigation 
that was further explored recently in Ref.~\cite{2023daSilva} for binary coupled
systems. With QT being understood as a complex dynamics of quantized 
vortices being  reconnected, it is also noticeable the recent experimental study 
on turbulent motion in quantum fluids, reported in Ref.~\cite{2023Makinen}, in which the 
authors consider a regime in quantum fluids where the role of vortex 
reconnections for turbulence can be neglected. From theoretical side, there is 
also a recent numerical simulation in~\cite{2020Muller}, considering 
Kolmogorov and Kelvin wave cascades within a generalized 
model for quantum turbulence, that includes beyond mean-field corrections.
Further, on the progress and status of QT investigations in BECs, 
we highlight the following reports~\cite{Tsatsos2016,Madeira2020b}, in which
more related references can be found.

Nevertheless, it is worthwhile to point out that most of the above mentioned studies 
 on QT are relying only upon non-dipolar atomic cases. 
Besides the actual increasing interest in vortices being generated in quantum dipolar gases
\cite{Lahaye2009,2012Kawaguchi,Mulkerin2013,2019Prasad,2021Prasad}, as well as
all other BEC investigations using dipolar atomic samples~\cite{2022Klaus,Chomaz2023},
the possibility of quantum turbulence in dipolar BECs remains almost unexplored,
except from a previous study in \cite{2018Bland} characterizing turbulence in the context of a dipolar
Bose gas condensing from a highly non-equilibrium thermal state, without external forcing.
Therefore, we understand as timely to investigate the possible occurrence of QT in 
a dipolar BEC submitted to a stirring mechanism. 
As detailed in the next sections, the occurrence of turbulence is characterized in our study by 
assuming defined regions of atom-atom parameters, as the repulsive contact $s-$wave
and dipole-dipole interactions. However, our analyses are not limited to the given   
values and can be extended to a larger range of parameters. For that, we consider a
stirring circular mechanism to initially produce the dynamics leading to vortex-pair
production, which eventually produces turbulence in the condensed quantum fluid.
Further, in order to bring out the impact of the DDI, we vary the strength of the DDI
by tuning the dipole angle defining the direction of the external magnetic field, as will be
shown. Thus, we hope this work can be helpful 
for experimental realizations of QT in dipolar BECs. 

The paper is structured as follows: In Sect.~\ref{sec2}, together with
our notation, we set the basic GP stirring model formalism with dipole-dipole 
interactions, with Sect.~\ref{sec3} furnishing the main results related to the vortex
nucleation considering two stirring models. In Sect.~\ref{sec4}, following a 
detailed analysis on the vortex dynamics through the kinetic energy spectra, we 
provide some evidences of quantum turbulence occurring when vortex-antivortex pairs
start to be nucleated. 
Finally, Sect.~\ref{sec5} presents our final considerations and conclusions.

\section{Gross-Pitaevskii stirring model with dipole-dipole interaction}
\label{sec2}
In the mean-field approximation of a dilute dipolar BEC of atoms with mass $m$, 
assuming two-body contact interactions between the $N$ atoms, the following 
three-dimensional (3D) time-dependent Gross-Pitaevskii (GP) equation is obtained 
for the wave function $\Psi({\mathbf r},t)$ normalized to one
\cite{Lahaye2009,Baranov2012}: 
{\small
\begin{align}
{\rm i}\hbar\frac{\partial \Psi({\mathbf r},t)}{\partial t}& =
\left(-\frac{\hbar^2}{2m}\nabla^2+ V({\mathbf r},t) + g_{3D}
 \left\vert \Psi({\mathbf r},t)\right\vert^2 \right)\Psi({\mathbf r},t)
 \notag \\ 
 &+N \int d^3{\mathbf r}' U_{\mathrm{dd}}({\mathbf  r}-{\mathbf r}')
 \left\vert\Psi({\mathbf r}',t)\right\vert^2 
 \Psi({\mathbf r},t), \label{eqn:dgpe}
\end{align}
}where $g_{3D}\equiv 4\pi\hbar^2 a_sN/m$ (with $a_s$ being the atom-atom $s-$wave
scattering length) is the cubic nonlinear contact parameter. 
$U_{dd}({\mathbf  r}-{\mathbf r}')$ provides the nonlocal long-range dipole-dipole
interaction between atoms at distances $|{\mathbf  r}-{\mathbf r}'|$, with
$V({\mathbf r},t)$ being an external pancake-like confining harmonic trap
supplemented by a time-dependent stirring interaction, to be detailed in this 
section. As considering the time-dependent mean-field formalism \eqref{eqn:dgpe} 
with possible vorticities, a convenient defined parameter related to the $N-$body
density is the {\it healing length},  
given by the inverse of the square-root of the chemical potential $\mu$ 
($\xi\equiv \hbar/\sqrt{m\mu}$), obtained by equating the quantum pressure and the interaction energy. 

\subsubsection*{Dipole-dipole interaction}
The long-range interaction between dipolar atoms with magnetic moments $\mu_A$, 
located at {$\bf r$} and {$\bf r'$}, when considering the tunability of the magnetic
dipolar interaction in quantum gases, is detailed in Ref.~\cite{Giovanazzi2002}.
With direction defined by an angle $\alpha$, it is applied an external magnetic field 
${\bf B}(t)$, which is a combination  of a static component $(B\cos\alpha) \hat{z}$
in the $z-$direction, with a fast-rotating (with frequency $\Omega$) transversal 
component $B\sin\alpha[\cos(\Omega t)\hat{x}+\sin(\Omega t)\hat{y}]$, 
such that within a cycle the atoms can be considered as remaining near the same position.
Given this condition, an average of the interaction can be performed
within a period, which provides an extra factor $(3\cos^2\alpha-1)/2$ 
multiplying the original dipole-dipole potential aligned along 
the $z-$direction. This procedure results in the cylindrically symmetric DDI
{\small\begin{eqnarray}
U_{\mathrm{dd}}({\bf r -r'})=
\frac{\mu_0 \mu_A^2}{4\pi}\frac{(1-3\cos^2 \theta_d)
}{\vert{\bf r -r'}\vert^3}\left(\frac{3\cos^2\alpha-1}{2}\right)
,\label{Udd}
\end{eqnarray}
}where $\mu_0$ is the permeability of free space, $\theta_d$ is the angle 
between the $z-$axis and the vector position of the dipoles ${\bf r -r'}$.
The angle $\alpha$ is defining the inclination of the dipole moments
$\boldsymbol{\mu}_A$ relative to the $z-$direction. In our pancake-like 
confinement (with most of atoms close to the transversal plane), 
the ${\bf r -r'}$ can be assumed in a plane perpendicular to the $z-$direction,
with $\theta_d\approx 90^\circ$. 
Therefore, the angle $\alpha$ turns out to be the key parameter in Eq.~\ref{Udd} 
to manipulate and alter effectively the DDI (independently on the $\mu_A$
values), from repulsive (when $\alpha<\alpha_M$) to attractive 
(when $\alpha>\alpha_M$) interactions, where $\alpha_M\approx 54.7^\circ$ is 
known as the {\it magic angle} (when the DDI is reduced to zero).
In our numerical approach, we consider the $^{168}$Er as the sample atom in 
the choice we made for the magnetic moment $\mu_A$ that gives the maximum 
repulsive value of the DDI strength (when $\alpha=0$), 
with $\mu_A=7\mu_B$ (where $\mu_B$ is the Bohr magneton). 
By changing the dipolar atom, 
together with the respective masses, the maximum DDI strength has to be re-adjusted 
correspondingly. The more recently dipolar atom produced in a BEC experiment, 
the $^{151}$Eu, has about the same value ($\mu_A=7\mu_B$) as $^{168}$Er, with  
$^{164}$Dy and $^{52}$Cr, having $\mu_A=10\mu_B$ and $6\mu_B$, respectively.

\subsubsection*{Confining potential perturbed by moving circular obstacle}
The confining trap potential $V({\mathbf r},t)$ is defined within a model 
that contains a pancake-like cylindrically symmetric harmonic interaction, 
with longitudinal and radial frequencies, $\omega_z$ and $\omega_\rho$, 
respectively, and large aspect ratio $\lambda=(\omega_z/\omega_\rho)^2
\sim 100$, perturbed in the transversal direction by a time-dependent 
penetrable Gaussian-shaped potential $V_G(x,y,t)$, expressed by
\begin{eqnarray}
V({\mathbf r},t)\equiv \frac{m\omega_\rho^2}{2}\left[(x^2+y^2)+
\lambda z^2\right]+V_G(x,y,t),
\label{trap}
\end{eqnarray}
where 
{\small\begin{eqnarray}\label{VG}
V_{G}(x,y,t)& \equiv &A(t) \exp\left(-\frac{\left[x-x_0(t)\right]^2+
\left[y-y_0(t)\right]^2}{2\sigma^2}\right)
\end{eqnarray}
}is modeling a possible experimental realization with a stirring mechanism that 
uses a laser-detuned 2D obstacle moving circularly within a fixed radius $r_0$ 
and given frequency $\nu$. In the above Gaussian distributions, 
$x_0(t)\equiv r_0 \cos(\nu \,t)$ and $y_0(t) \equiv r_0 \sin(\nu \,t)$ 
are giving the instant position of the obstacle in the 2D plane, 
with $\sigma$ being the corresponding standard radial deviation
(close to half-width of the distribution).
The amplitude (strength) of the perturbation, $A(t)$, is represented by
{\small\begin{equation}
A(t) \equiv  A_{0}\left[1+\varepsilon \sin(\omega_A\,t)\right]
= A_{0}\left[1+\varepsilon \sin\left(\frac{\omega_A}{\nu}\,
\nu t\right)\right],\label{At}\end{equation}
}within two variant types, with respect to the time dependence: 
For type-I, $A(t)=A_0$ is invariant ($\varepsilon=0$); 
and, for type-II, $A(t)$ vibrates with frequency $\omega_A$ and displacement 
factor $\varepsilon\ne 0$. 
In both cases, $A_0$ is assumed to be close to 90\% of the stationary chemical
potential $\mu$.
In Fig.~\ref{fig01}, we have a pictorial representation of the pancake-like
condensate with the moving laser obstacle. 

Our approach to producing the dynamics relies on considering the Gaussian-shaped
time-dependent interaction \eqref{VG} in the GP formalism \eqref{gpe3d}, 
for a condensed dipolar system with the DDI \eqref{Udd} having the strength 
controlled by the angle $\alpha$. Therefore, once the trap aspect
ratio $\lambda$ and the atom-atom two-body scattering length  are fixed, 
besides the DDI parameter $\alpha$, the other main parameters are 
the ones of the Gaussian model \eqref{VG}; namely, the amplitude $A(t)$ 
(in the two variant types), width $\sigma$, position $r_0$, and angular 
frequency $\nu$. 
The choice of the parameters, such as $r_0$ and the two-body repulsive 
interaction $a_s$, were fixed after some preliminary investigation leading to 
the production of vortex pairs. Within this purpose, having the repulsive contact
interaction driving the original size of the condensate, the radial position 
should be kept not close to the center (where the circular speed will come out 
being unrealistic high for vortex production), as well as not in the very-low
density region (where the expected increasing vortex numbers occur in a
non-uniform region of the quantum fluid). 
We are aware that, by considering $a_s$ and $r_0$ fixed, the
parameter $\alpha$ must be restricted within some limits to keep repulsive 
the total interaction (sum of contact and dipolar), within a condensate with
radius enough larger than $r_0$, for the analysis of the vortex dynamics.

\begin{figure}[!ht]
\vspace{-0.2cm}
\centering\includegraphics[width=0.99\linewidth]{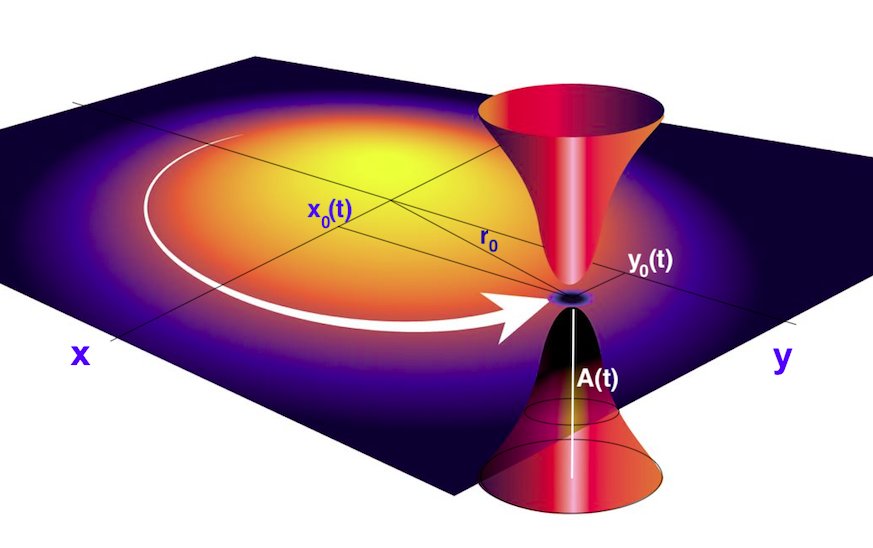}
\vspace{-1.cm}
\caption{(color online) 
Graphical representation, in arbitrary units,  
of a bell-shaped laser stirring {\it penetrable} obstacle, with
Gaussian format defined by~\eqref{VG} and \eqref{At}], circularly moving 
inside a BEC fluid with velocity $r_0\nu$ and radius $r_0$.}
\label{fig01}
\end{figure}

In the remaining part of this section, we first express the 3D GP equation 
(\ref{eqn:dgpe}) in dimensionless form, following by a reduction of the 
formalism to 2D, which is validated by a large aspect ratio $\lambda$.
By keeping our notation close to the one used in Ref.~\cite{Kumar2017} (for 
binary BEC with DDI), with the frequency and length units given by $
\omega_\rho$ and $\ell_\rho\equiv \sqrt{\hbar/(m \omega_\rho)}$, the 
full-dimensional variables are replaced by the corresponding dimensionless 
ones, as
${\bf r}/\ell_\rho \to {\bf r}$; $ \omega_{\rho} t\to t$;
$a_s/\ell_\rho\to a_s$; 
$\ell_\rho^{3/2}\Psi\to \Psi$. Also, by factoring the energy unit
$\hbar\omega_\rho$, the trap interaction \eqref{trap} and the stirring potential
\eqref{VG} will remain having the same formal expression, which means  
$V({\bf r},t)/({\hbar\omega_\rho})\to V({\bf r},t)$  and
$A(t)/({\hbar\omega_\rho})\to A(t)$.
Therefore, in dimensionless quantities, (\ref{eqn:dgpe})  is replaced by
{\small
\begin{eqnarray}
{\rm i} \frac{\partial \Psi({\mathbf r},t)}{\partial t}& = 
\Big(-\frac{1}{2}\nabla^2+V({\mathbf r},t) +g \vert {\Psi({\mathbf{r}},{t})}
\vert^2 \Big) \Psi({\mathbf r},t) \\ \nonumber 
&+g_{\mathrm{dd}}\left[\int d^3{\mathbf{r}}'{\displaystyle 
\frac{3\cos^2\alpha-1}{2 \vert{\bf r -r'} \vert^3}} \vert \Psi({\mathbf{r}}',t)
\vert^2 \right]\Psi({\mathbf r},t),
\label{gpe3d} 
\end{eqnarray}
}where $g\equiv g_{3D}/ (\hbar\omega_\rho \ell_\rho^3)$
and $g_{\mathrm{dd}}\equiv N\mu_0 \mu_A^2/(4\pi \hbar\omega_\rho \ell_\rho^3)$.
In terms of a defined dipole length $a_{dd}\equiv \mu_0\mu^2/(12\pi\hbar\omega\ell_\rho^2)$,
we can write $g_{dd}$ as $g_{dd}=3N(a_{dd}/\ell_\rho)$.
The stability of a dipolar BEC depends on 
the external trap geometry, \textit{e.g.}, a dipolar BEC is stable or unstable 
depending on whether the trap is pancake- or cigar-shaped, respectively. The
instability usually can be overcome by applying a strong pancake trap with 
repulsive two-body contact interaction. The external trap helps to
stabilize the dipolar BEC by imprinting anisotropy to the density 
distribution. So, with the dynamics of the dipolar BEC strongly confined in the
axial direction ($\lambda \gg 1$), $\Psi({\bf r},t)$ can be decoupled as 
{\small \begin{align}\Psi({\bf r},t)\equiv 
\chi(z)\psi({\bf {\boldsymbol\rho}},t)\equiv
\left(
\frac{\lambda}{\pi}\right)^{{1}/{4}}
\exp\left(\frac{-\lambda z^2}{2}\right) \psi({\bf {\boldsymbol\rho}},t),\nonumber 
\end{align} }where $\boldsymbol{\rho}\equiv(x,y)
\equiv(\rho\cos\theta,\rho\sin\theta)$, allowing the dynamics in the 
$z-$direction be integrated out. This procedure is straightforward for non-dipolar
BEC systems. However, we still need to perform the 2D reduction of the DDI 
configuration-space term, which is followed by the double $z-$integration. 
Apart of the DDI parameter $g_{dd}$, the 2D expression for the DDI is given 
by~\cite{2012Wilson}
\begin{equation}
V^{(d)}(\boldsymbol{\rho }-\boldsymbol{\rho}^\prime)
=\int dz dz^\prime|\chi(z)|^2 |\chi(z^\prime)|^2 
\frac{3\cos^2\alpha-1}{2|{\bf r}-{\bf r}^\prime|^3}.
\label{2D-DDI}\end{equation}
Given the corresponding Fourier transforms to the momentum space
$({\bf k}_\rho,k_z)$, for this
potential, being $\widetilde{{V}}^{(d)}({\bf k_\rho})$, for the 
axial density $\widetilde{n}({k_z})$, 
as well as for the 2D densities,
${\widetilde{n}}({\bf k_\rho})$,  
we have the  following identification~\cite{Kumar2019}:
{\small\begin{equation}
\int d\boldsymbol{\rho }^\prime V^{(d)}(\boldsymbol{\rho }-
\boldsymbol{\rho}^\prime)|\psi^\prime|^{2}
=\mathcal{F}_{2D}^{-1}\left[ \widetilde{{V}}^{(d)}({\bf k_\rho}){\widetilde{n}}
({\bf k_\rho})\right].
\end{equation}
}In the 2D momentum space, the DDI can be expressed as the combination of two 
terms, considering the orientations of the dipoles $\alpha$ and projection of the
Fourier transformed in momentum space. One term is perpendicular, with the other 
parallel to the direction of the dipole inclinations~\cite{2012Wilson,Zhang2016}, 
expressed by
\begin{eqnarray}
h_{2d}^{\perp}(\mathbf{k}_{\rho})
&=&
2-3\sqrt{\frac{\pi}{2\lambda}}k_\rho
\exp{\left(\frac{k_{\rho }^{2}}{2\lambda}\right)} 
{\rm erfc}\left( \frac{k_{\rho }}{\sqrt{2\lambda}}\right), 
\\
h_{2d}^{\parallel}(\mathbf{k}_{\rho})
&=&
-1+3\sqrt{\frac{\pi}{2\lambda}}\frac{k_x^2}{k_\rho}
\exp{\left(\frac{k_{\rho }^{2}}{2\lambda}\right)} 
{\rm erfc}\left( \frac{k_{\rho }}{\sqrt{2\lambda}}\right), 
\label{hparallel}
\end{eqnarray}
with ${\rm erfc}(x)$ being the complementary error function of $x$. 
For the parallel term, the projection of the polarizing field 
onto the $x-y$ plane is assumed in the $x-$direction.
by considering that all directions $\theta_k$ are possible, and 
$k_x=k_\rho\cos\theta_k$ in (\ref{hparallel}), 
for a polarization field rotating in the $(k_x,k_y)$ plane, we can average 
$k_\rho^2\cos^2\theta_k$, replacing this term by 
$k_\rho^2/2$. By combining the two terms according 
to the dipole orientations $\alpha$, the total 2D momentum-space DDI,
$\widetilde{{V}}^{(d)}({\bf k_\rho})=
\cos^2(\alpha) h_{2d}^{\perp}(\mathbf{k}_{\rho})
+ \sin^2(\alpha) h_{2d}^{\parallel}(\mathbf{k}_{\rho})$,
becomes proportional to  
$h_{2d}^{\perp}(\mathbf{k}_{\rho})$:
\begin{equation}
{\widetilde{V}}^{(d)}({\bf k_\rho}) = \frac{3\cos^2\alpha-1}{2}
 h_{2d}^{\perp}(\mathbf{k}_{\rho})
\label{DDI-2D}.
\end{equation} 
Therefore, with the dipolar term in the Fourier-transformed momentum space, we obtain the effective 
2D equation for the pancake-shaped dipolar BEC as
\begin{eqnarray}
 {\rm i}\frac{\partial \psi}{\partial t}&=&\left\{
 -\frac{1}{2}\nabla_{\rho}^2+\frac{\rho^2}{2}+ V_{G} +g_{s}|\psi|^2 \right.
\nonumber\\
&+&\left. g_{dd}\int \frac{d^2k_{\rho}}{4\pi^2}
e^{i \mathbf{k}_{\rho}.\tilde{\rho}}\tilde{n}(\mathbf k_{\rho})
\widetilde{{V}}^{(d)}({\bf k_\rho}) \right\} \psi,
\label{gpe_scaled}
\end{eqnarray}
where $\psi\equiv\psi(\boldsymbol{\rho}, t)$ is the 2D wave function, normalized 
to one, $g_s\equiv \sqrt{8\pi\lambda} {a_s N}/{\ell_\rho}$, and $V_G\equiv V_{G}
(\boldsymbol{\rho},t)$ given by \eqref{VG}. In these dimensionless units, for a
circular moving obstacle, we can conveniently write $V_G$ as
{\small\begin{eqnarray}\label{VG2}
V_{G}(\boldsymbol{\rho},t)& = &A(t)\exp\left[-\frac{\rho^2+r_0^2-2\rho\,r_0
\cos(\theta-\nu t)}{2\sigma^2}\right].
\end{eqnarray}
}From this expression, we notice that the period $T$ for a complete one-loop
circular movement (potential returning to the same value) is given by $T=2\pi/\nu$,
such that we could replace $\nu t$ by $2\pi t/T$, with $t$ given in terms of the
loop-period $T$. For the solution of the above Eq.~(\ref{gpe_scaled}), in order to
investigate the vortex nucleation and dynamics of the vortex pair productions in
the dipolar BECs, we need to combine the usual split-step Crank-Nicholson method
with the Fast Fourier Transform approach~\cite{Kumar2015}. In our numerics, we 
have used a $256 \times 256$ grid size with $\Delta x=\Delta y=0.1$ for both $x$
and $y$ (units $\ell_\rho$), with time step $\Delta t=0.0001 \omega_\rho^{-1}$. 
Further, we have confirmed that the results are not affected by doubling the
aforementioned grid sizes and grid spacing. Along with our study, we also kept
fixed the following parameters related to the atom-atom 
interactions and stirring Gaussian model:
$a_s=50\,a_0$, $a_{dd}=66\,a_0$, $N=1.5 \times 10^4$, $A_0=36\hbar\omega_\rho$, 
$r_0=3.5\ell_\rho$, and $\sigma=1.5\ell_\rho$, where $a_0=5.29177\times 10^{-11} m$ 
is the Bohr radius and $\ell_\rho=1\times10^{-6}m$. The choice of these parameters 
is motivated by possible experimental realization considering the dipolar $^{168}$Er,
with two-body scattering length enough repulsive such that the effect of
the dipolar interaction can better be evaluated by changing the tilting
DDI angle $\alpha$. 
The radial position of the obstacle ($r_0$) inside the condensate was arbitrarily chosen at an 
approximate average distance between the center and the border limits of the condensate,
considering that the critical velocities for the production of vortex-pairs are given by  
$v_c=r_0\nu_c$.
By varying the $r_0$ position inside the main part of the condensed fluid, one can verify
that no vortex pairs can be produced when $r_0\to 0$.

\section{Vortex nucleation by stirring Gaussian potential}
\label{sec3}
This section is concerned with the dynamical production of vortex dipoles and vortex 
clusters by the time-dependent stirring interaction. To obtain a stable ground state 
density profile, in our simulations the Eq.~(\ref{gpe_scaled}) is first solved by using 
imaginary time $(t \rightarrow {-it})$, without the Gaussian obstacle ($A(t)=0$). 
Next, this solution is evolved in real-time with the obstacle ($A(t)\neq 0$).
Once considered fixed the shape of the Gaussian model for the obstacle (amplitude and 
width), with its radial position inside the condensate, the principal model parameters  
are the strength of the DDI and the angular frequency $\nu$ of the obstacle, implying 
in a corresponding linear velocity $v=\nu r_0$. We have considered two types of simulations, 
with the type-II  ($\varepsilon \neq 0$) differing from type-I ($\varepsilon = 0$) by 
additionally verifying the effect of vibration (with frequency $\omega_A$) in the amplitude 
of the obstacle.

\subsection{Critical velocities: Phase diagrams}\label{sec3A}
With the circular movement of the obstacle fixed at a constant radial position
$r_0$, the numerical simulation starts (at $t=0$) with a linear ramping of the 
time-dependent amplitude $A(t)$ of the stirring interaction, from $A(t)=0$ to 
$A(t)=A_0$, within the time interval $\Delta T=0.0025\omega_\rho^{-1}=25\Delta t$. 
Along the simulation, this value $A_0$ remains fixed in case of type-I model, 
being time vibrating with frequency $\omega_A$ in case of type-II model. 
The obstacle inside the condensate is assumed moving with a constant frequency $\nu$, 
Once considered the model interaction parameters, as the two-body contact $a_s$ 
and dipole-dipole strength (provide by $a_{dd}$ and the angle $\alpha$, our
results for the two defined stirring models, related to the production of
vortex-antivortex pairs and vortex clusters, are resumed in the two panels 
shown in Fig.~\ref{fig02}, in which we are providing diagrams for the 
DDI angles $\alpha$ as functions of the rotation frequency $\nu$ ($v=r_0\nu$). 
The panels (a) and (b) are, respectively, for non-breathing ($\varepsilon=0$ ) 
and breathing ($\varepsilon=0.4$ with $\omega_A=2\omega_\rho$) modes of the Gaussian obstacle. 
Also, in the time-dependent expression of the amplitude \eqref{At}, we assume 
$A_0=36\hbar\omega_\rho$ (which is about $90\%$ of the chemical potential $\mu$) and 
$\sigma=1.5\ell_\rho$. Considering such moving penetrable obstacles, the vortices are 
created in the form of dipoles consisting of two vortices with opposite circulations, 
within a dynamical process qualitatively different from the typical hard cylinder 
case, where vortices can also be generated individually due to the vortex 
pinning effect in the density-depleted region~\cite{Lim2022}. 
Within this process, for fixed values of $\alpha$, critical velocities
can be verified for the production of vortex-antivortex pairs, or vortex 
clusters, defining three 
regions: with no vortex (A), with vortex-antivortex pairs emerging in 
distinguishable time-gap intervals (B), and with vortex pairs
emerging as clusters (in indistinguishable time intervals, almost  
simultaneously). In the panels of Fig.~\ref{fig02}, considering the given 
fixed parameters, the separation between regions A and B is provided by 
solid-thick lines, that interpolate precise numerical results (showing by 
stars in case of type-I model). 
By increasing even more the frequency $\nu$, another critical border 
transition can be approximately verified, when the vortex pairs 
start emerging as clusters. 
Panel (a) shows more explicitly how the vortex pairs produced in the first cycle 
($N_{v\bar{v}}$) change by increasing $\nu$. 
The first thick-solid line ($N_{v\bar{v}}=1$, for $\alpha$ versus $\nu$)
is interpolating exact critical points obtained in our calculation, represented
by stars. The other long-dashed lines are guiding eyes for the 
verified crescent $N_{v\bar{v}}$ as $\alpha$ increase with $\nu$. 

The dynamical process of the vortex nucleation can be understood as follows:
When the Gaussian obstacle moves faster than a critical velocity $v_c$, 
energy is transferred into the BEC density by changing the superfluid velocity 
field near the obstacle. As the accumulated energy $E$ exceeds a 
certain threshold $E_v$, the energy will dissipate via vortex-antivortex 
emission~\cite{Lim2022}, represented by the thick-solid line between 
regions A and B, as described above; or via vortex-antivortex clusters for
even larger velocities of the obstacle.
 Further, a remark has to be done related to the
observed number of pairs $N_{v\bar{v}}$ in our simulations: 
once the vortex dipoles are produced in the first cycle, by keeping the same 
rotational speed the number of vortices produced in the next cycles is clearly
reduced, which can be explained by the changes in the fluid dynamics as the 
time flows. The production of vortex pairs occurs more easily with the obstacle
rotating inside an originally stationary uniform fluid.
As the time flows, the fluid near the obstacle is not anymore in the same 
stationary state as it happens when starting the first loop.  For different 
combinations of dipolar and contact interactions, a more detailed 
analytical investigation on this dynamics is being under consideration in a
following related work, in which the plan is to investigate the interaction between
the produced vortex pairs and their persistency inside the fluid.

\begin{figure}[!ht]
\centering\includegraphics[width=1.0\linewidth]{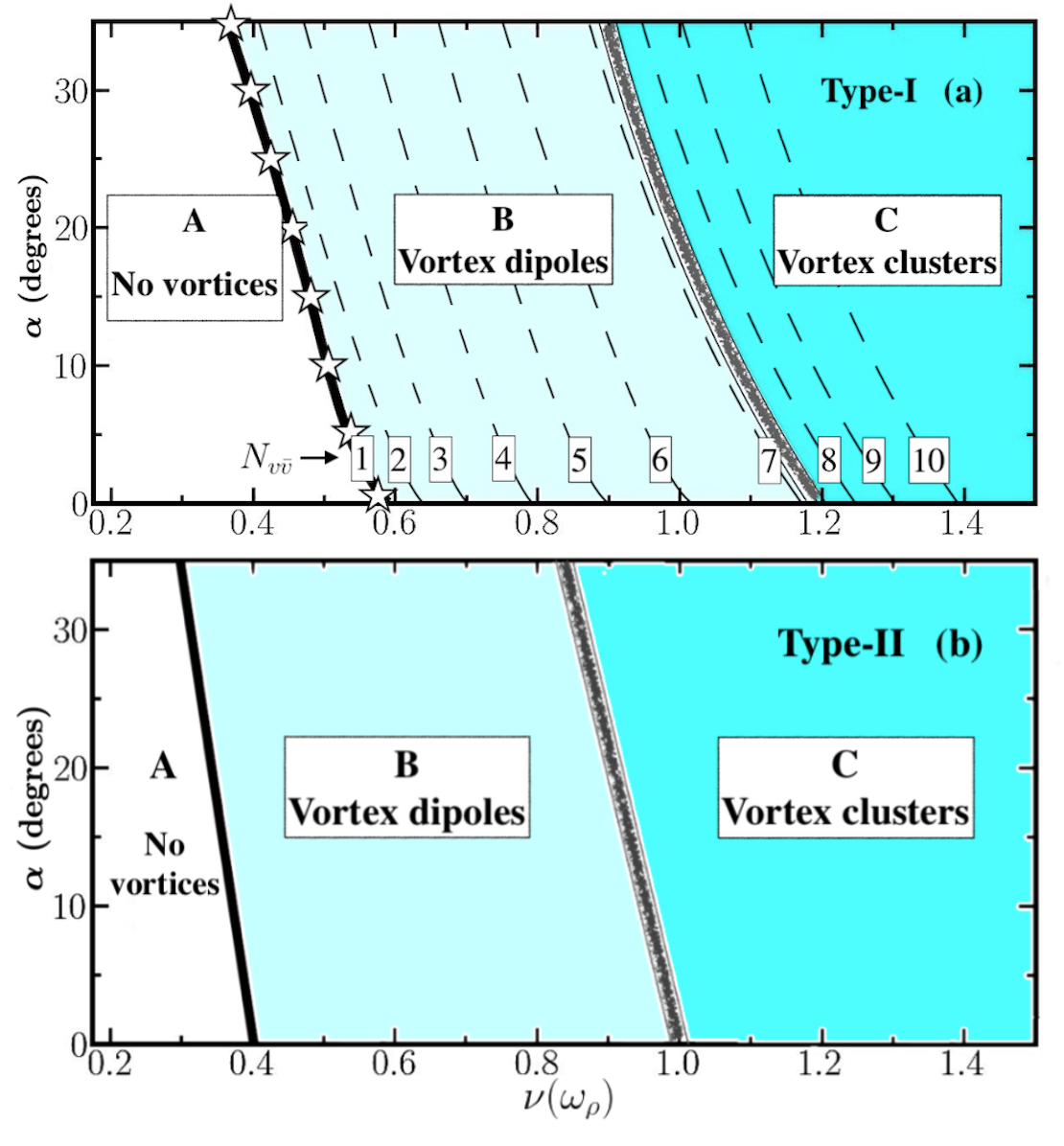}
\caption{(color online) 
 Diagrams for $\alpha$ (angles for DDI strengths) versus 
$\nu$ (rotational frequencies)
for type-I (a) and type-II (b) models, described by 
\eqref{VG}, with $\varepsilon=0$ (a) and $(\varepsilon,\omega_A)=
(0.4,2\omega_\rho)$  (b) (with $A_0=36\hbar\omega_\rho$ and 
$\sigma=1.5\ell_\rho$ in both cases). 
The critical rotational frequencies for emergence of the first vortex pair are
given by the thick solid lines separating regions A and B. 
With type-II following a behavior similar as type-I, panel (a)
shows more explicitly how the critical $\nu$ for $N_{v\bar{v}}=1$
is obtained by interpolating exact results (indicated by stars).
The long-dashed lines are for increasing number of vortex pairs, as indicated.
The separation between regions B (vortex dipoles) and C (vortex clusters) 
is represented by a blurred band line in both panels. Other common parameters
are:  $r_0=3.5\ell_\rho$ (obstacle location), 
$a_s=50\,a_0$, $a_{dd}=66\,a_0$, and $N=1.5 \times 10^4$.}
\label{fig02}
\end{figure}

\begin{figure}[!ht]
\centering\includegraphics[width=1.0\linewidth]{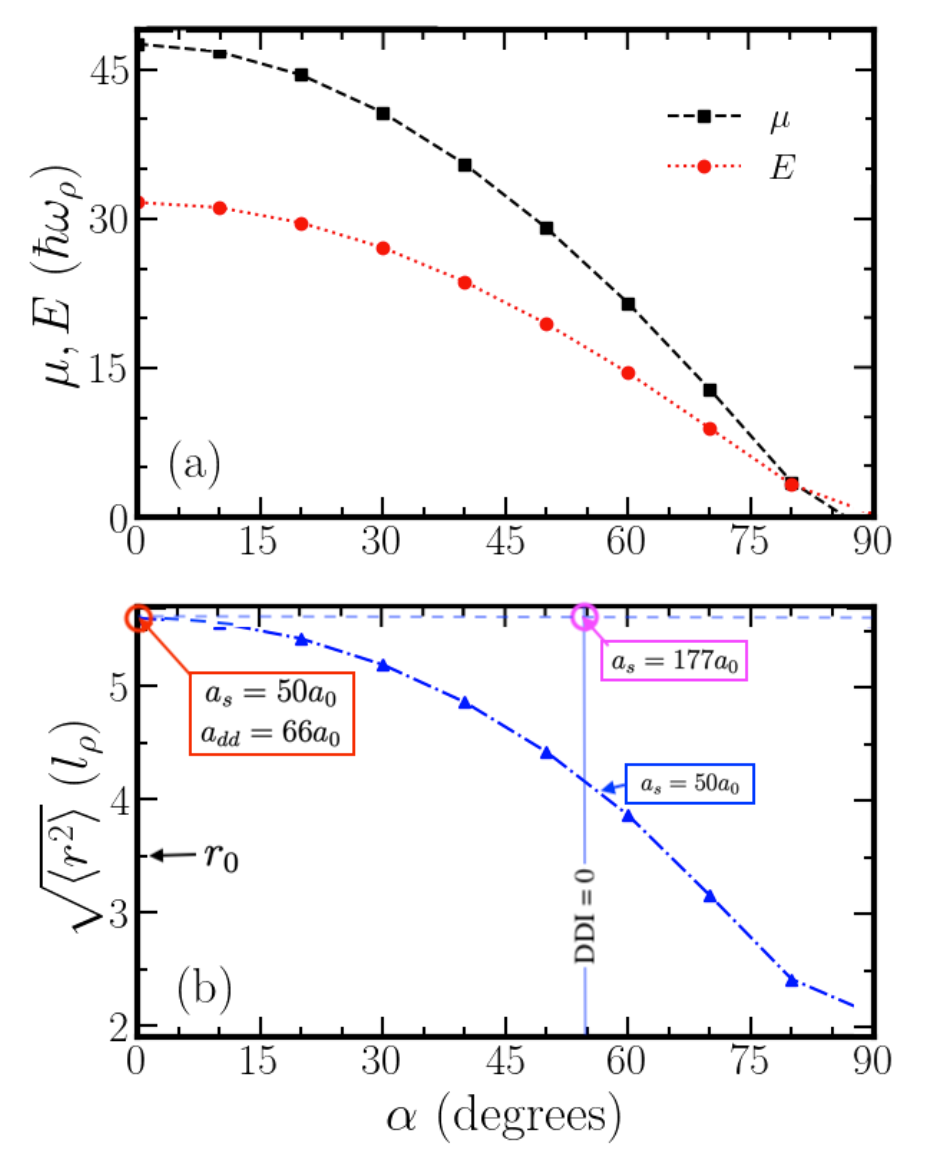}
\caption{(color online)
Effect of DDI ($\alpha$) variations on 
the ground-state chemical potential ($\mu$) and energy ($E$) 
are shown in (a), with the corresponding rms radius
($\sqrt{\langle r^2\rangle}$) in (b).
For $\alpha< 54.7^\circ$ the DDI 
is repulsive, whereas for $\alpha> 54.7^\circ$ it is attractive. The two 
indicated circles in (b) refer to the necessary change in the contact 
interaction $a_s$ to keep the respective condensed clouds with about the same size.
}
\label{fig03}
\end{figure}

By examining in more detail the type-I model [panel (a) of Fig.~\ref{fig02}], 
when the repulsive DDI is at the maximum ($\alpha=0$),
the minimum critical frequency to produce one pair in the first cycle
($N_{v\bar v}=1$) is verified being $\nu_c\approx 0.57\omega_\rho$, with 
the number of pairs increasing almost linearly up to 
$\nu\approx1.2\omega_\rho$, when the vortex-cluster productions start to occur 
(with more than one pair at each instant of time).
For the type-II case, as shown in panel (b) of Fig.~\ref{fig02}, the critical 
frequency to producing vortex dipoles at $\alpha=0^\circ$ is $\nu_{c}\approx 
0.4\omega_\rho$, with $N_{v\bar{v}}$ increasing with $\nu$ up to $\approx 1\omega_\rho$
when the pairs start being produced as clusters.
The shifted region borders to lower values of $\nu$ (by going from type-I to type-II models) 
are consistent with the extra dynamics due to the amplitude vibration.
By reducing the repulsive DDI, with increasing $\alpha'$s, in both cases $\nu_c$ is consistently 
reduced, as shown for $\alpha$ going from 0$^\circ$ to 35$^\circ$.
In fact, this behavior is expected considering the other parameters that are kept fixed.
With the condensate size being reduced (as the contact interaction remains at $a_s=50a_0$), 
the radial position of the moving obstacle,  fixed at $r_0=3.5\ell_\rho$, end up being located in 
relatively less-dense region of the condensate.

In all given numerical simulations represented in Fig.~\ref{fig02}, 
except for the Gaussian amplitude and DDI angle, the other parameters 
are kept fixed, with the trap aspect ratio $\lambda=100$, 
interaction strengths ($a_{dd}=66a_0$ and $a_s=50a_0$) and the
number of atoms $N=1.5\times 10^4$, as indicated in the caption of
Fig.~\ref{fig02}. Here, we should remind that the assumed magnetic moment,
$\mu_A=7\mu_B$, implies $^{168}$Er or $^{151}$Eu (with the
corresponding change in the mass unit) as the atomic sample, 
such that the angle $\alpha$ must be correspondingly shifted
when considering another dipolar species (as $^{164}$Dy or $^{52}$Cr).
The critical velocities ($r_0\nu_c$) are verified being 
approximately constant for reasonable changes of $r_0$ inside the condensate, 
such that no vortex pairs are produced near $r_0=0$.

The results shown in Fig.~\ref{fig03}, for the chemical potential $\mu$ and total energy $E$, 
[panel (a)], with the corresponding root-mean-square (rms) radius [panel (b)], 
as functions of the DDI angle $\alpha$, without the Gaussian obstacle,
are helpful to estimate the more convenient radial position for the obstacle in the
following simulations with the obstacle. As verified, by inspecting both panels (a) and (b),
the rate of changing for $\mu$ is approximately the same as the rate of changing for 
$\langle r^2\rangle$.
As the position of the obstacle approaches the center, the frequency $\nu$ 
needs to increase, such that for $r_0\to 0$ no vortex can be created by 
the stirring. Therefore, in principle, one could conclude that 
larger values of $r_0$ will be more favorable for the production 
of vortex dipoles or vortex clusters, implying the obstacle located in the
low-density region of the trap. However, this possible choice will bring us
the role of the other constraints we have to consider, such as the contact 
atom-atom interaction.
By assuming fixed and repulsive $a_s=50a_0$, the DDI parameter 
$\alpha$ is controlling the size (and corresponding the chemical potential) of the condensate. 
So, we need to keep the DDI enough repulsive ($\alpha<\alpha_M$), otherwise with 
$\alpha>\alpha_M$ the stirring position will be located in the too-low density region, 
or even outside the condensate, as one can verify by considering $\alpha\ge 70^\circ$, when
the rms radius of the condensate becomes smaller than $r_0$.
With these concerns, restricted by the parameter regions in our study of vorticity, 
$\alpha$ was chosen enough smaller than the magic angle $54.7^\circ$.

As to keep the study of the dynamics in similar conditions, considering the size of
the condensed cloud, with $\mu$ and $\langle r^2\rangle$ not deviating more than about
20\% from their maximum values by varying the interaction parameters (contact and DDI), 
we have verified as appropriate to keep the DDI limited to the repulsive region, with 
$\alpha$ between $0^\circ$ and 35$^\circ$. 
This, obviously, restricted by the other model parameters that are kept fixed 
in our study, such as the contact interaction and position of the obstacle.
By going to attractive DDI region, in order to maintain the size of the condensate 
in stable conditions, as well as not too small considering the position of the 
obstacle, one needs necessarily to increase the repulsive contact interaction,
such as going to $a_s= 177a_0$, as shown in Fig.~\ref{fig03}(b). 
As an alternative, for a smaller stable cloud, one could reduce  
the radial position of the obstacle.
Following our vortex dynamics study, in the next subsection, a 
comparative analysis is also provided in section III.B for two condensates with equivalent
rms and $\mu$, with the DDI of one of the condensates reduced to zero.

\begin{figure*}[!ht]
\centering\includegraphics[width=18cm,height=8.5cm]{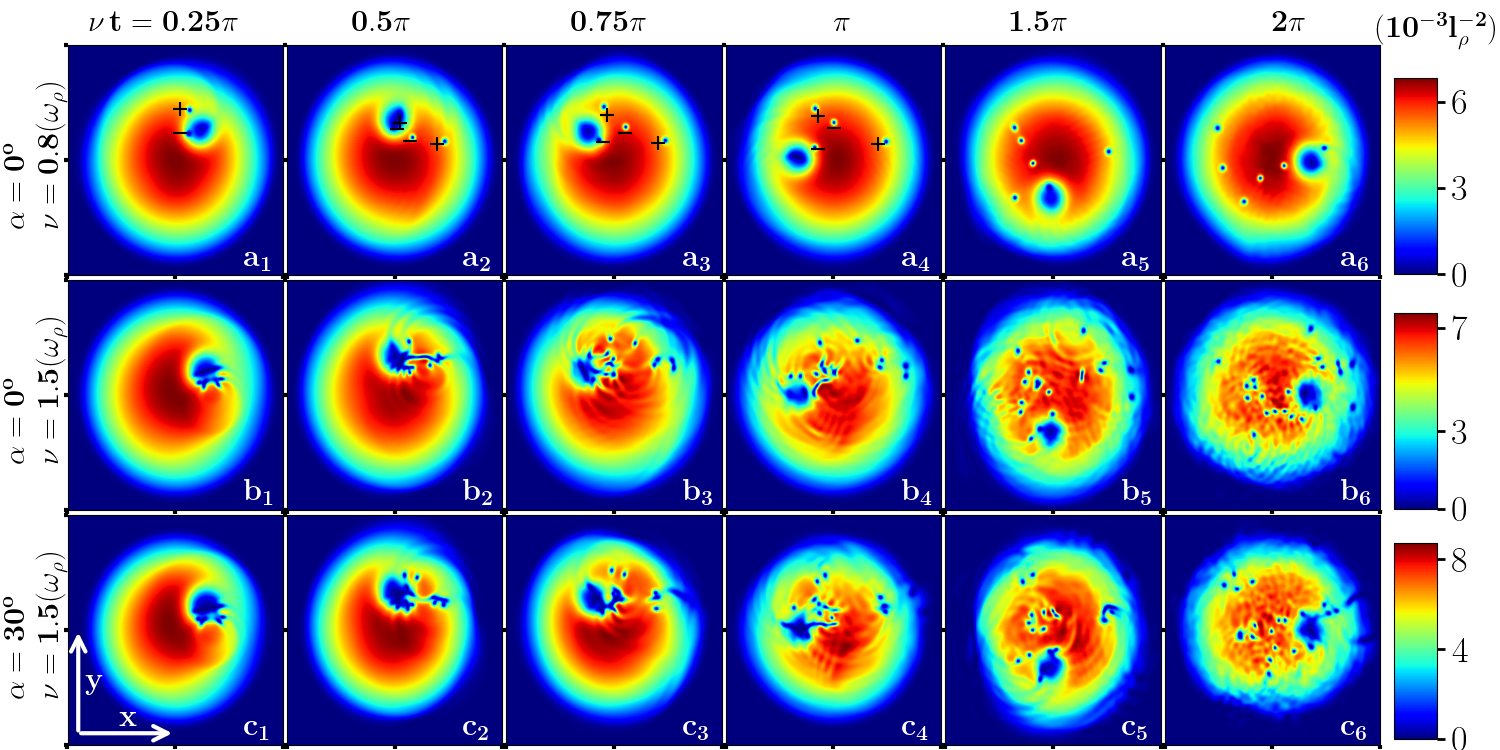}
\caption{(color online) 
For the type-I ($\varepsilon = 0$) rotating obstacle at $r_0=3.5$, this 
figure shows time snapshots of full-numerical results for the 2D
densities (levels indicated by color bars), with the production of 
vortex-antivortex (upper row) and vortex-cluster (middle and lower rows), 
in the first cycle, as indicated: $\nu t=\pi/4$ (a$_{1}$,b$_{1}$,c$_{1}$) till 
$\nu t=2\pi$ (a$_{6}$,b$_{6}$,c$_{6}$). The rotation frequencies and
DDI angles are, respectively, $(\nu,\alpha)=(0.8\omega_\rho,0^\circ)$ (top row), 
$=(1.5\omega_\rho,0^\circ)$ (middle row), and $=(1.5\omega_\rho,30^\circ)$ (bottom row).
Other fixed parameters (not explicitly indicated) are the same as 
the ones given in the caption of Fig.~\ref{fig02}. The vortex $(-)$ and
antivortex $(+)$ signs are indicated in the first four panels of the 
upper row, respectively.}
\label{fig04}
\end{figure*}
\begin{figure}[!ht]
\vspace{-1cm}
\includegraphics[width=1\linewidth]{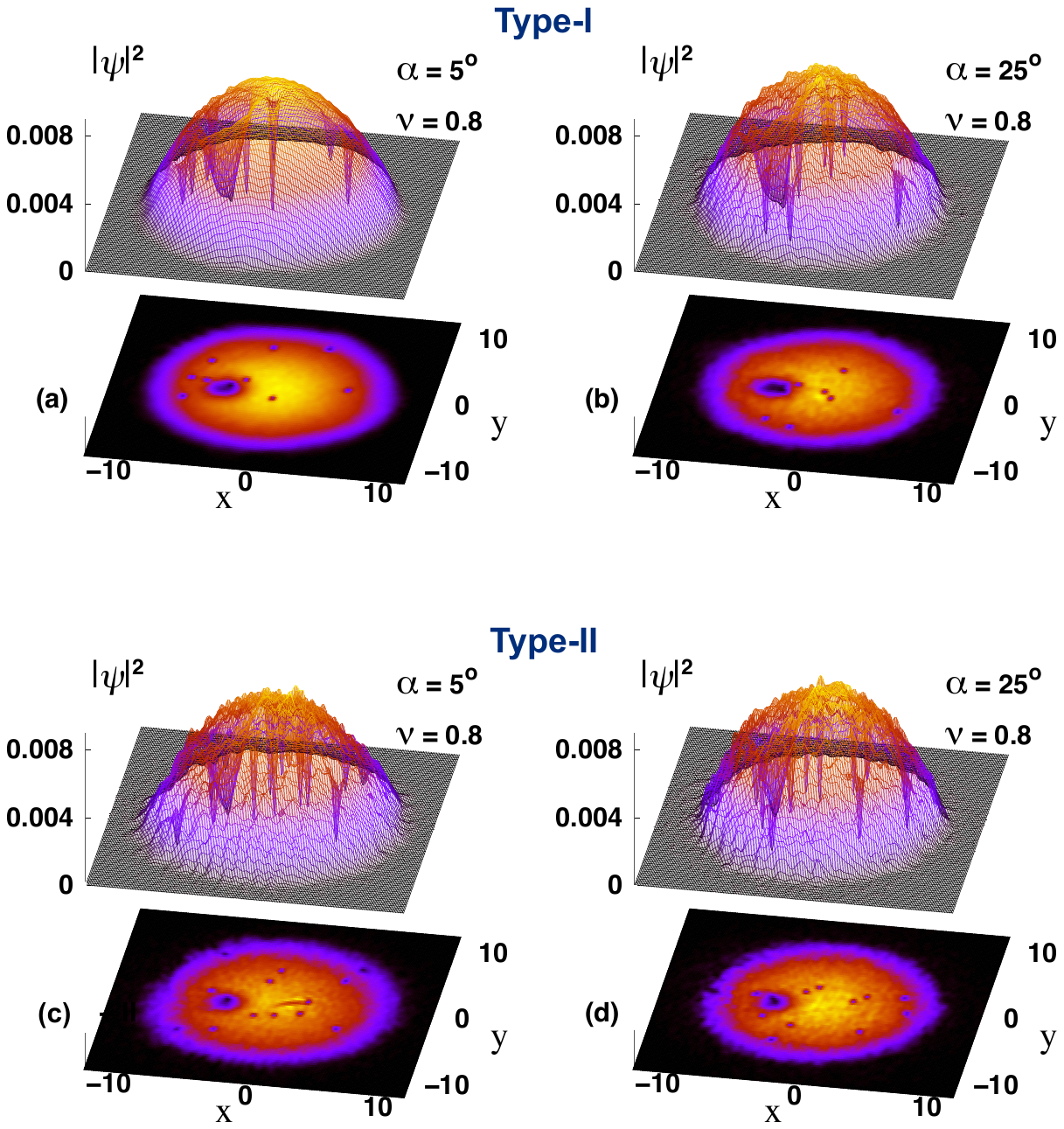}
\vspace{-1.5cm}
\caption{(color online)
Results obtained for the 2D densities ($|\psi|^2$, in $l_\rho^{-2}$ units), given 
in 3D plots, with corresponding projections in the $x$ and $y$ plane (units $\ell_\rho$),
for type-I [$\varepsilon=0$, panels (a) and (b)] and type-II 
[$(\omega_A=2\omega_\rho,\varepsilon=0.4$, panels (c) and (d)] models. We consider two sets 
of DDI angles $\alpha=5^\circ$ [panels (a) and (c)] and $25^\circ$ [panels (b) and (d)], with 
fixed rotational frequency ($\nu=0.8 \omega_\rho$), as indicated.
They are representing the vortex-antivortex production in region B (shown in Fig.~\ref{fig02}), 
with identical arbitrary time such that $\nu t$ is slightly larger than $3\pi$. The other 
parameters are, respectively, the same as given in the caption of Fig.~\ref{fig02}.}
\label{fig05}
\end{figure}

\subsection{Vortex dynamics within the condensate}\label{sec3B}

In this section, we select some specific parameters for which we can verify more clearly 
the associated dynamics in the formation of the vortex-antivortex pairs, by following the 
evolution of the densities. The differences between the two types of stirring approaches 
rely only on the additional dynamics introduced by a periodic variation in the amplitude g
of the stirring interaction, given by \eqref{At}. In view of that, we can first resume 
our main results for the case in which the amplitude is kept constant, followed by an 
analysis of the type-II case (when the amplitude is vibrating).
For this purpose, we present Fig.~\ref{fig04} considering the type-I model, 
immediately followed by the corresponding discussion for the type-II model, in order to 
verify possible additional effects (if any) introduced by the periodic vibration of the 
stirring amplitude.

Figure~\ref{fig04} shows three rows of snap-shot density plots obtained from 
full-numerical calculations considering the type-I Gaussian model for the stirring. 
These results illustrate the nucleation of the vortex dipoles and vortex clusters 
during the movement of the non-breathing penetrable Gaussian obstacle with two 
different frequencies ($\nu=0.8\omega_\rho$ and $1.5\omega_\rho$) and two DDI angles 
($\alpha=0^\circ$ and 30$^\circ$).  The Gaussian obstacle with its position is represented 
in each of the panels by the hollow (minimum density) moving anti-clockwise inside the fluid. 
In the upper row of panels, for $\nu=0.8\omega_\rho$, which is picked from region B in 
Fig.\ref{fig02}(a), the Gaussian obstacle nucleates the vortices in the form of 
vortex dipoles. The nucleation starts close to $t=1\omega_\rho^{-1}$ [or $\nu t=\pi/4$, as verified 
in panel (a$_1$)].  In the other two rows, for $\nu=1.5\omega_\rho$, which are picked from region 
C in Fig.\ref{fig02}(a), the Gaussian obstacle nucleates the vortices in the form of 
clusters. In these cases, with two different DDI angles, the nucleation starts before, 
close to $t=0.5\omega_\rho^{-1}$ [or $\nu t=\pi/4$, as verified in panels (b$_1$) and (c$_1$)].
 
In order to appreciate how the vortex-antivortex, as well as clusters of vortex,
are emerging, we are showing six snapshots for the first cycle, with $\nu t$
going from $\pi/4$ up to $2\pi$.
As verified in the first row, with $\alpha=0^\circ$ and $\nu=0.8\omega_\rho$, we have the 
production of about three vortex-antivortex pairs within a cycle.  
For the stirring model, a constant amplitude $A_0=36\hbar\omega_\rho$ is assumed in this case. 
More specifically, the first row [$\nu=0.8\omega_\rho$, with $\alpha=0^\circ$] 
corresponds to region B of the upper panel of Fig.~\ref{fig02}, with the 
vortex-antivortex pairs being produced regularly at different time intervals, 
as the obstacle moves anti-clockwise around the circle.
The second and third rows refer to the vortex-cluster production region C
in the upper panel of Fig.~\ref{fig02}, with $\nu=1.5\omega_\rho$ and $\alpha=0^\circ$
(second row) and $30^\circ$ (third row), when vortex clusters are being
produced (more than one pair at each time interval).
Corresponding to $\nu=0.8\omega_\rho$ and $1.5\omega_\rho$, the respective stirring velocities are 
$v=2.8$ and $v=5.25$. They are selected in correspondence with the 
results previously shown in the upper panel of Fig.~\ref{fig02}.
The effect of the rotation $\nu$ can be seen by comparing the first with the 
second row, as $\nu$ is changed to $1.5\omega_\rho$ with $\alpha$ having the same value.
Similarly, to see the effect of $\alpha$, we consider the case with $\nu=1.5\omega_\rho$,
with the second and third row varying $\alpha$ from 0$^\circ$ to $30^\circ$. 
In this case, the difference is quite visible in the contour plots of the
densities, as well as the dynamics of the emerged vortex and antivortex. 
Reflecting the fact that the repulsive DDI for $\alpha=0^\circ$ is 
stronger than for $\alpha=30^\circ$, we can observe that the vortex-antivortex pairs
repeal each other more strongly in case that $\alpha=0^\circ$ [compare, for example, 
the positions of the emerged vortices in (b$_4$) with the ones in (c$_4$)].

One should notice that, differently from the case in which the obstacle moves 
linearly inside the condensate~\cite{Sabari2018}, the production of the 
vortex-antivortex pairs occurs with each vortex of the pair emerging at a 
slightly different time. This can be understood considering that the density 
distribution of the BEC fluid around the obstacle is not the same in both sides.
Within a counter-clockwise stirring rotation, as the fluid is denser in the 
internal left side, the emerging vortex [indicated by ($-$) in panels (a$_{1}$-a$_{4}$) 
of Fig.~\ref{fig04}] takes a slightly longer time to emerge than the associated 
antivortex [indicated by ($+$)]. 

Considering the type-II model, according to the results 
previously pointed out in Fig.~\ref{fig02} for the critical rotational
velocities necessary for vortex-antivortex productions,  
due to the additional dynamics introduced by the amplitude
vibration, more vortex pairs are verified emerging in a cycle than 
the ones observed in Fig.~\ref{fig04}. These results can 
already be verified by examining the two diagrams shown in 
Fig.~\ref{fig02}. For the type-II case, considering $A_0=36\hbar\omega_\rho$ and 
$\varepsilon=0.4$, the amplitude is vibrating from $A=21.6$ 
[when $\sin(\omega_A t)=-1$] till $A=50.4$ [when $\sin(\omega_A t)=1$]. 
With $\omega_A=2\omega_\rho$ the oscillating period is $5/2$ times the stirring cycle 
frequency, when $\nu=0.8\omega_\rho$; and $4/3$ in case $\nu=1.5\omega_\rho$.
Therefore, the production of vortex pairs and vortex clusters occurs at shorter time intervals
in the case of the type-II model, as the quantum fluid is more affected by the Gaussian 
amplitude vibration.

By verifying that the results for the densities are not significantly
different in both cases, in Fig.~\ref{fig05} we have selected 
parameters considering the region B of Fig.\ref{fig02}, with rotational 
frequency $\nu=0.8\omega_\rho$, presenting our results in a 3D illustrative format,
for the dynamics observed without (type-I) and with (type-II) amplitude vibration. 
The choice of the two DDI angles have the purpose of verifying how the dynamical 
behavior of the produced vortex pairs is changed by going  
from a more repulsive $\alpha=5^\circ$ to a less repulsive $\alpha=25^\circ$ condition.
The panels (a) and (b) refer to the type-I (non-breathing mode, $\varepsilon=0$), 
with panels (c) and (d) referring to the type-II [breathing mode, with ($\omega_A,\varepsilon)
=(2\omega_\rho,0.4)$]. In both cases, we select identical time snap-shots, with $\nu t$ slightly 
larger than $3\pi$ (second loop). 
By comparing the left with the right panels, in both cases, the obstacles (radially 
fixed at $r_0$) are moving inside regions of the respective condensates that have relatively
slightly different densities: It is in a denser region in case $\alpha=5^\circ$, because
the radius is larger than for $\alpha=25^\circ$. In other words, by keeping fixed the
radius $r_0$, to study the dynamics within similar conditions we need to 
restrict the values of the DDI to reasonable not too large angles, such that the obstacle
remains moving in similar density regions of the condensate. By taking larger values of 
$\alpha$ the system will be more attractive (unless, to compensate, we change another
interaction parameter, as the scattering length) with the obstacle position moving at 
too low density region (as close to the radial border limits) of the condensate.
As comparing with Fig.~\ref{fig04}, where we choose $\alpha=0^\circ$ and $\alpha=30^\circ$,
the motivation in changing slightly the angles, as shown in Fig.~\ref{fig05}, is to  
appreciate how the vortices are emerging by slightly decreasing the DDI (with $\alpha$ going
from 0$\circ$ to 5$^\circ$) or slightly increasing (with $\alpha$ going
from 30$^\circ$ to 25$^\circ$). 

\begin{figure}[!ht]
\centering\includegraphics[width=1\linewidth]{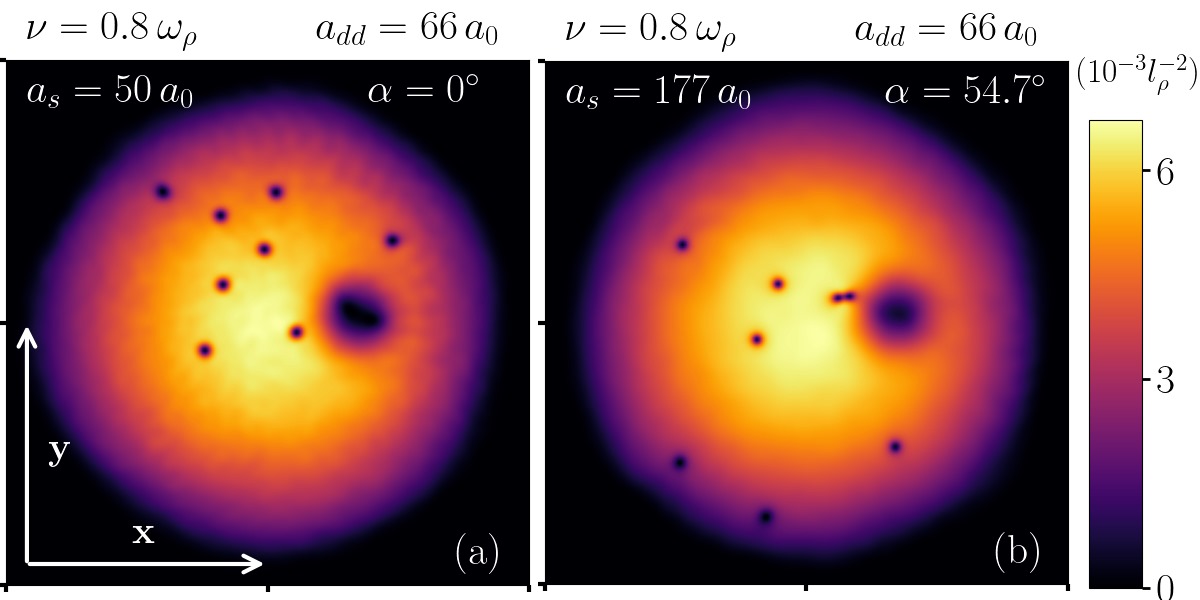}
\caption{
Snap-shots of densities obtained with maximum DDI ($\alpha=0$, at left) and
without DDI ($\alpha=\alpha_M=54.7^\circ$, at right) obtained near two cycles ($4\pi$).
At right $a_s=177a_0$, in order to have both clouds with about the same
rms radius. A corresponding gif-movie animation is presented in a Supplemental 
Material.
}
\label{fig06}
\end{figure}

\subsection{Role of DDI in the dynamics}
For a comparative study on the role of DDI for the vorticity and
turbulence, we need to consider two condensates such that one
of them is without DDI. For that, by starting with a simulation
in which $a_s=50a_0$, the maximum DDI is obtained at $\alpha=0$,
as indicated by the red circle inside Fig.~\ref{fig03}(b). As
to compare with this case, in the other simulation we
consider $\alpha$ fixed at the magic angle, implying the DDI is
zero. But, as we need to consider condensates with about
the same radial extension, such that the obstacle position will be 
not in the low-density region, we have to assume a larger value
$a_s=177a_0$ in order to have both condensates with about the 
same sizes. This second case is indicated by a magenta circle inside 
Fig.~\ref{fig03}(b). Resulting from this comparative simulation,
we have two snap-shots in Fig.~\ref{fig06},
which are taken when the rotation finish two cycles (4$\pi$). By 
looking to the dynamics from $t=0$, in both cases (with and without DDI)
we observe that the initial production of vortices occurs in a very similar way 
(slightly faster in case of DDI), implying that the critical velocities
verified in Fig.~\ref{fig02} are mainly due to the changes in chemical potential 
and rms radius. However, after the first cycle it is also visible in the dynamics that
the case with $\alpha=0$ (maximum DDI) presents stronger fluctuations 
in the cloud, absent for non-dipolar case, 
reflecting the different characteristics of the interactions in the fluid.
 In fact, a more close inspection by varying the rotation speed $\nu$ of the obstacle, 
considering two extreme cases with equivalent condensed cloud sizes, one in which the DDI is zero
(pure contact interactions with $a_s=177a_0$), with the other in which the contact interaction is zero 
(pure DDI, with $a_{dd}=91a_0$ and $\alpha=0$), it was verified that the production of 
vortex-antivortex requires smaller rotation speed in case of pure DDI.

In a Supplemental Material~\cite{DDIvsContact-animation}, we include explicitly our
results obtained for the dynamics, with $\nu=0.8\omega_\rho$, through an animation in 
which we can observe the time evolution of both condensed systems in the process of
vortex-antivortex emission, with ($\alpha=0$)
and without ($\alpha=\alpha_M$) DDI, from $\nu t=0$ till $\nu t=2\pi$. When removing
the DDI, the contact interaction has to be redefined (from  $a_s=50a_0$ to
$a_s=177a_0$) in order to maintain both condensates having about the same size, with
the obstacle kept fixed at $r_0=3.5\ell_\rho$.}
Still, the similarity of the two cases when started the dynamics
with the obstacle is indicating that the kind of interactions (contact or DDI)
are not so relevant for the initial formation of the vortex-antivortex pairs in 
the condensate, provided that both densities are initially found  
in the same conditions, having about the same sizes.
As the number of vortex pairs being created as time flows is drastically 
reduced, within this comparison, one can draw the conclusion that 
the vorticity and turbulence are mainly affected by the size of the 
condensed cloud, instead of the two kind of interactions we are 
considering (DDI or contact).
However, apart of the initial vortex production, in view of the different 
observed dynamics inside the fluid, a more dedicated conclusive investigation 
may be required.
As the results we are reporting rely only on the 
numerical solution of the corresponding circular-stirred GP formalism with 
two-body contact and dipolar interactions, without the assumption of possible beyond 
mean-field corrections, a more prospective analysis may not be difficult, but  
outside the scope of the present work.

\section{Spectral dynamics and quantum turbulence in stirred BEC with DDI}
\label{sec4}
Our aim in this section is to characterize the possible emergence of turbulent behavior in the 
dipolar condensate, by full-numerical investigation of the evolution and behavior of the
vortices within a spectral analysis. 
The results of our study are illustrated by selecting two different rotational frequencies of the 
obstacle (considering the regions B and C in the two panels of Fig.~\ref{fig02}), with repulsive DDI strengths
being typified by two distinct angles $\alpha$. In particular, the choice to keep the DDI enough
repulsive was determined by the other choices for the model parameters, such as the contact
interaction and radial position of the circularly moving obstacle.

Towards an understanding of possible quantum turbulence in a quantum
fluid as a cold-atom system described by the mean-field GP theory, one of the main 
characterization to look for is the Kolmogorov's classical scaling law $k^{-5/3}$  
in the incompressible kinetic energy spectrum~\cite{Kolmogorov}, as pointed out in 
several works and reviews in this direction~\cite{Barenghi2001,Tsatsos2016,Madeira2020b}. 
As reported already in 1997 in Ref.~\cite{Nore1997}, considering nonlinear
Schr\"odinger equation 
solutions with possible implications in experiments for Helium superfluid, it was
found that low-temperature superfluid turbulence follows approximately Kolmogorov's
scaling. The vorticity dynamics of the superflow were shown to be similar to that
of the viscous flow, including vortex reconnection. Directly connected with 
cold-atom experiments in the last 15 years, vorticity and QT was reported in
oscillating BECs in Refs.~\cite{Henn2009,Henn2010}. In nonuniform BEC the
occurrence of QT was investigated numerically in Ref.~\cite{Horng2009}.
The actual increasing interest in the QT investigations in cold-atom physics can be
traced from Refs.~\cite{White2010,Seman2011,Reeves2012,Kwon2014,White2014,Fujimoto2015,Cidrim2016,
Kobayashi2021,Reeves2022}, having as strong motivating factor possible links 
between QT and its classical counterpart.
Therefore, by assuming the Kolmogorov scaling behavior of the kinetic energy
spectrum as a parameter for a universal description of turbulence, one has to
characterize the length scales by considering the necessary stationary states with
enough number of vortices being produced.
For the analysis of vorticity and the occurrence of turbulence in an ensemble of
particles, the relevant quantity is the associated kinetic energy, which can be
decoupled in two terms, considering the compressibility. The key concepts concerned
with the energy spectra of vortex distributions in 2D QT, together with a 
discussion on similarities and differences with 2D classical turbulence has been
explored for homogeneous compressible superfluid in Ref.~\cite{2012Bradley}. For 
our present analysis of the energy spectra, in the next we follow some details provided more 
recently in Refs.~\cite{2022Bradley,2023daSilva}. Of particular interest in our
case is the investigation of a dipolar confined BEC system, which is under external
laser stirring periodic perturbation.

\begin{figure}[!ht]
\centering
\includegraphics[width=1.0\linewidth]{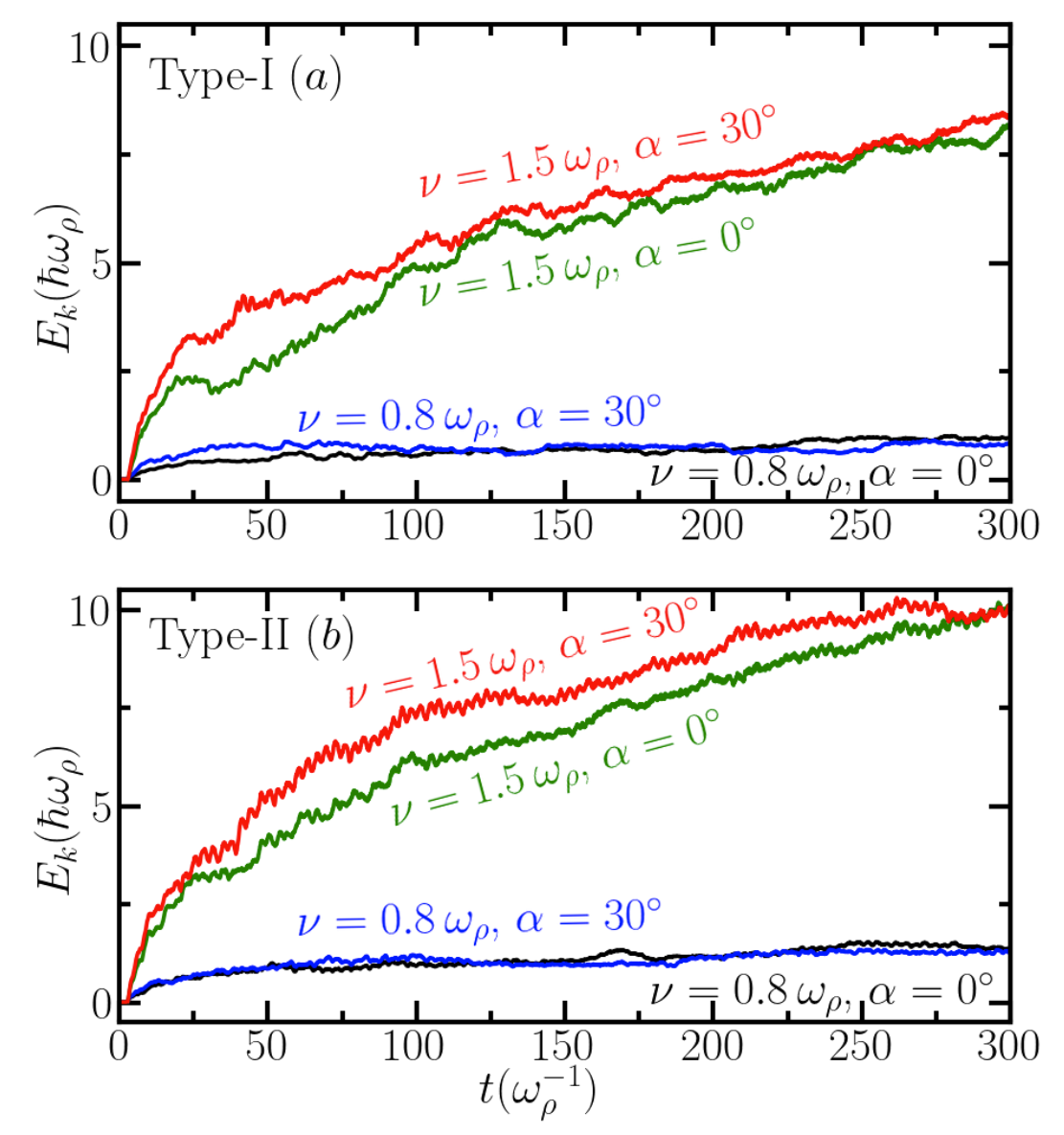}
\caption{(Color online) Time evolutions of the total kinetic energies for the 
type-I [panel (a)] and type-II [panel (b)]stirring models.
In both the cases, the results are for two sets of rotational velocities ($\nu=0.8\omega_\rho$ 
$1.5\omega_\rho$) and 
DDI $\alpha$ angles
($\alpha= 0^\circ$ and $30^\circ$), as indicated close to
the respective lines. }
\label{fig07}
\end{figure}

\begin{figure}
\centering\includegraphics[width=1.0\linewidth]{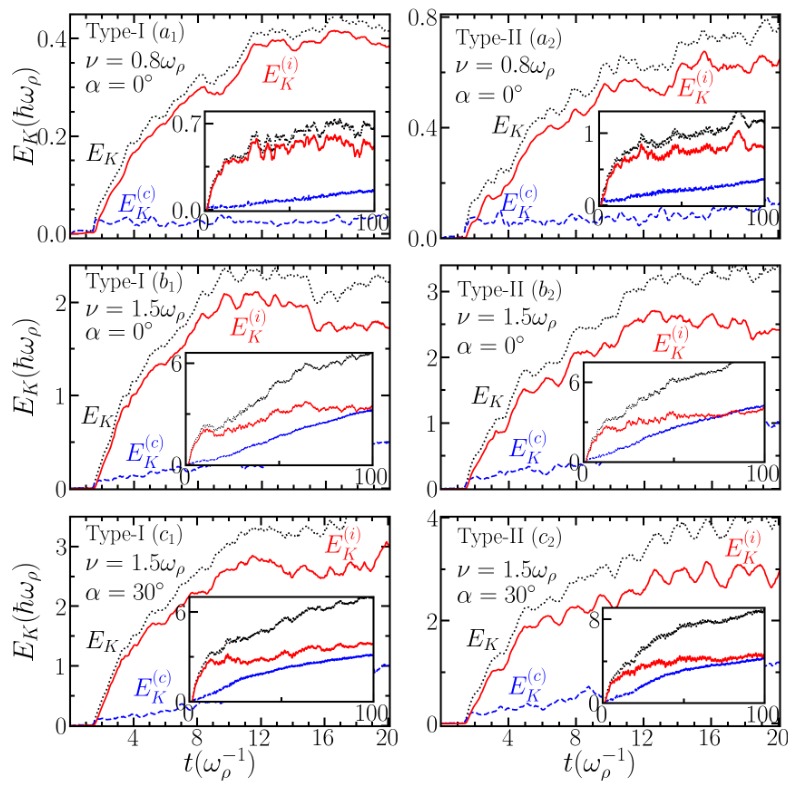}
\caption{(Color online) Time evolution of the total ($E_K$, black-dotted lines), 
compressible ($E_K^{(c)}$, blue-dashed lines), and incompressible  
($E_K^{(i)}$, red-solid lines) kinetic energies (in $\hbar\omega_\rho$ units) for type-I (left panels) and 
type-II (right panels) stirring motion of the Gaussian obstacle. The rotational
frequencies $\nu (=0.8\omega_\rho, 1.5\omega_\rho)$ and DDI angles $\alpha (=0^\circ, 30^\circ)$ are
indicated inside the respective panels. The insets are shown the corresponding 
long-time behaviors. The time intervals, identified in the bottom panels, 
are common to all panels. 
}
\label{fig08}
\end{figure}

\begin{figure}[!ht]
\centering\includegraphics[width=1.\linewidth]{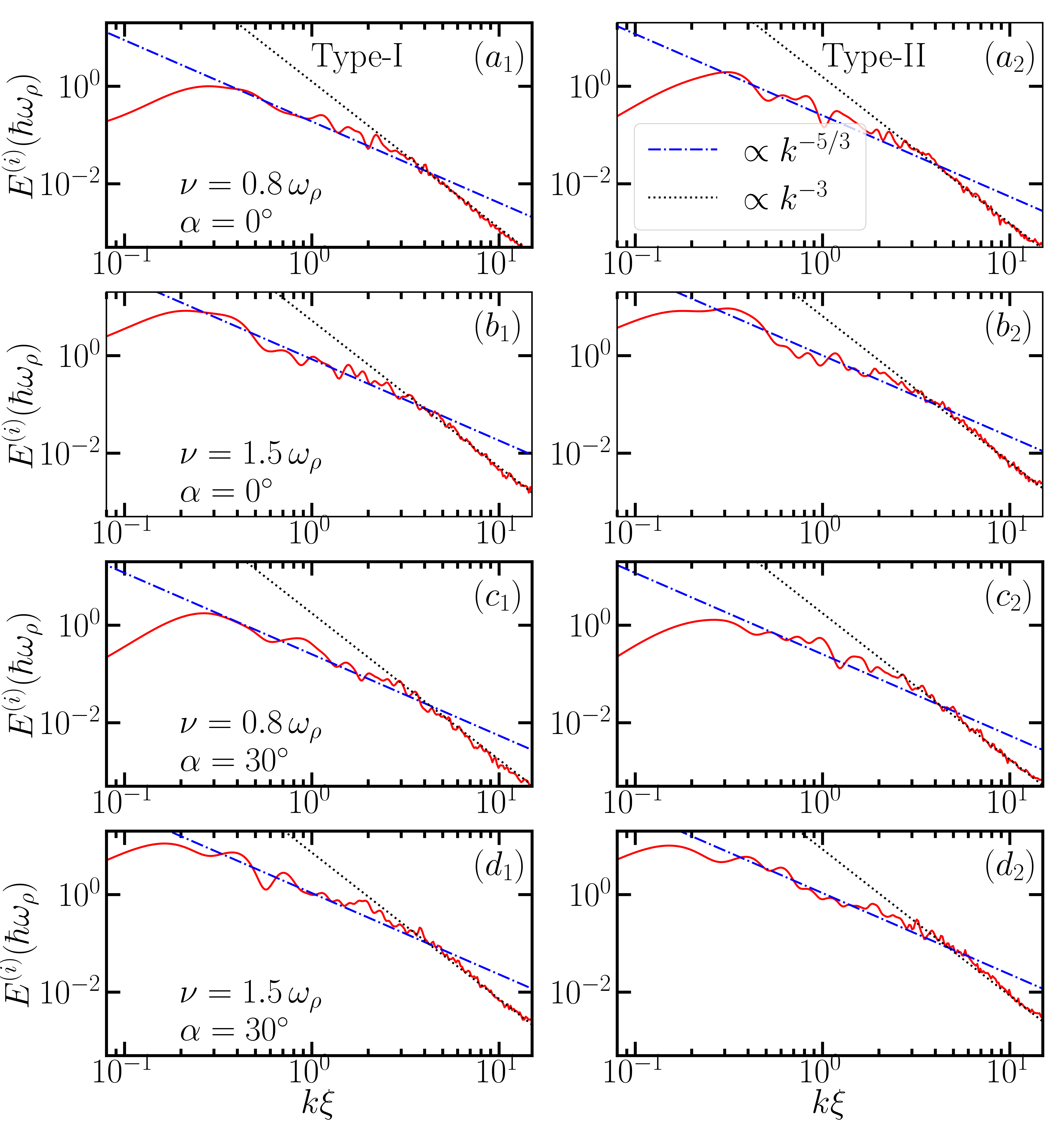}
\caption{
Incompressible kinetic energy spectra, $E^{(i)}(k)$, obtained for the
type-I [(a$_1$)-(d$_1$) panels] and the type-II [(a$_2$)-(d$_2$) 
panels] models, as functions of the dimensionless $k\xi$ (where
$\xi$ is the healing length).
The rotation frequencies $\nu$
(0.8$\omega_\rho$ and 1.5$\omega_\rho$) and 
DDI angles $\alpha$ ($0^\circ$ and 
$30^\circ$) are being indicated inside 
the respective panels. At each panel, the  
$E^{(i)}(k)$ results refer to averaging 
over 10 samples in the vortex emission
regimes identified in Fig.~\ref{fig08}
(left) (for type-I); and 
Fig.~\ref{fig08}(right) (for type-II).
The straight blue-dot-dashed and 
black-dotted lines are guidelines to follow
the respective $k^{-5/3}$ and $k^{-3}$ behaviors.}
\label{fig09}\end{figure}

\subsection{Kinetic energies: Compressible-incompressible decomposition}
\label{sec:energy}
From the GP Eq.~\eqref{gpe_scaled}, which  provides the time-dependent 
density solution $n(\boldsymbol{\rho},t)$, the total energy is 
{\small\begin{eqnarray}
E(t)&=&
\int d^2\boldsymbol{\rho}\left[\frac{1}{2}|\nabla_{\rho}\psi|^2 + \frac{\rho^2}{2}+ 
V_G(\boldsymbol{\rho},t)n(\boldsymbol{\rho},t)\right] 
\nonumber\\&+&\frac{g_{s}}{2} \int d^2\boldsymbol{\rho}\; n^2(\boldsymbol{\rho},t)
+E_{dd}
\label{En}, 
\end{eqnarray}
}where $E_{dd}$ refers to the DDI contribution. The total kinetic energy 
contribution, $E_K\equiv\int d^2\boldsymbol{\rho}\left[
\frac{1}{2}|\nabla_{\rho}\psi|^2\right]$, is the main term we are concerned with 
in this section. Its time evolution relies on the density contributions. 
By considering the related current density ${\bf j}(\boldsymbol{\rho},t)$ in terms
of the velocity field ${\bf v}(\boldsymbol{\rho},t)$, we have ${\bf j}
(\boldsymbol{\rho},t)=n(\boldsymbol{\rho},t){\bf v}(\boldsymbol{\rho},t)$, or 
\begin{equation}\label{current}
{\bf v}(\boldsymbol{\rho},t) = \frac{1}{2\mathrm{i}\; |\psi|^2} { 
\big[ \psi^\star \nabla_{\rho}\psi - \psi \nabla_{\rho} \psi^\star \big]}.
\end{equation}
The associated kinetic energy term, being expressed by
\begin{align}\label{Ken}
E_K(t) = \frac{1}{2} \int d^2{\boldsymbol{\rho}}\;n(\boldsymbol{\rho},t)\;
|\mathrm {\bf v}(\boldsymbol{\rho},t)|^2,
\end{align}
can be decomposed into compressible $E_K^{(c)}(t)$ and incompressible 
$E_K^{(i)}(t)$ parts. For that, we define the density-weighted velocity field 
${\bf u}(\boldsymbol{\rho},t)\, \equiv\,\sqrt{n(\boldsymbol{\rho},t)} 
{\bf v}(\boldsymbol{\rho},t)$, that can be split into an incompressible part, 
${\bf u}^{(i)}\equiv {\bf u}^{(i)}(\boldsymbol{\rho},t)$ satisfying 
$\nabla_{\rho} . {\bf u}^{(i)} = 0$, and a compressible one, ${\bf u}^{(c)}
\equiv {\bf u}^{(c)}(\boldsymbol{\rho},t)$, satisfying 
$\nabla_{\rho} \times{\bf u}^{(c)} = 0$. 
Therefore, with
${\bf u}(\boldsymbol{\rho},t)\, =\, {\bf u}^{(i)}(\boldsymbol{\rho},t) + 
{\bf u}^{(c)}(\boldsymbol{\rho},t)$, the two parts of the kinetic energy can be
defined by  
{\small
\begin{equation}\label{Ken-split}
\hspace{-0.2cm}E_K(t)=E_K^{(i)}(t)+E_K^{(c)}(t)
\equiv \frac{1}{2} \int d^2\boldsymbol{\rho}\; \left[|{\bf u}^{(i)}|^2 +
|{\bf u}^{(c)}|^2 
\right].\end{equation}
}
Associated with these energies, due to the time-dependent stirring interaction, 
an effective torque is experienced by the system, which can be obtained from 
the corresponding $z-$ component of the operator 
${\pmb\tau} = {\bf r}\times F=-{\bf r}\times \nabla V_G$, 
which in polar coordinates is reduced to
\begin{eqnarray}\label{torque}
{\bf \tau}_z (\boldsymbol{\rho},t)&=& 
-\frac{\partial}{\partial\theta}V_G(\boldsymbol{\rho},t). 
\end{eqnarray}

In correspondence with previously presented results, on the production of vortex
dipoles, vortex clusters, as well as the time evolution of the densities, by 
considering the above formalism for the total $E_K$, compressible $E^{(c)}_K$,
and incompressible $E^{(i)}_K$ kinetic energies, we are presenting some
sample results in the next, by considering some specific significant values 
of the parameters, guided by the previously obtained results reported 
in Fig.~\ref{fig02}. 
The long-time evolution of the total kinetic 
energy is first shown in two panels given in Figs.~\ref{fig07}. 
For that, we choose $\alpha=0^\circ$ and $\alpha=30^\circ$  for the angle
controlling the DDI strength, corresponding both to repulsive interactions
with maximum being at $\alpha=0^\circ$. For the stirring periodic frequency, 
we have assumed $\nu=0.8\omega_\rho$ and $\nu=1.5\omega_\rho$. The stirring is applied at a fixed 
distance given by $r_0=3.5$, implying velocities $v_0=$2.8 and 5.25, respectively. 
As shown in Fig.~\ref{fig02}, for the two types of dynamics
of the stirring, the case with $\nu=0.8\omega_\rho$ refers to the sector where we have 
vortex dipoles production; whereas with $\nu=1.5\omega_\rho$ being the region with 
vortex clusters production. Apart from the fact that type-II is more energetic 
than type-I, expected due to the stirring vibration in addition to the circular velocity, 
we observe that both types have similar general behavior as
related to the DDI, with type-II being more sensible to $\alpha$ in the long-time
evolution than type-I for higher velocities. However, when considering low 
velocities (vortex-dipole production region) both cases are almost unaffected by
the DDI strength. For both types, I and II, fast stabilization of the total 
energy is obtained with low speed.

\noindent With Fig.~\ref{fig07}, we are illustrating the time evolution of 
the total kinetic energies for the type-I stirring model, considering two values
for the frequency, and two values for the DDI, respectively given by 
$\nu=0.8\omega_\rho,\; 1.5\omega_\rho$ and $\alpha=0^\circ, 30^\circ$. We choose 
only this case to show the long-time
behavior of the kinetic energies when increasing the stirring velocities. Similar
behavior can be verified for the type-II model. As seen, the kinetic energy $E_K$
increases significantly when we increase the strength of $\nu$ from the 
vortex-dipole region ($\nu=0.8\omega_\rho$) to the vortex-cluster region $\nu=1.5\omega_\rho$. 
With respect to the DDI variation, measured by the parameter $\alpha$ (stronger DDI 
implying in $\alpha=0$), one can notice that the kinetic energy is reduced, as 
we increase the DDI, from  $\alpha=30^\circ$ to $\alpha=0$.   

In the next results we are going to discuss, we are more concerned with the 
not-large period of time, below $t=20\omega_\rho^{-1}$, in which we can associate turbulence
behaviors of the condensed fluid.
The time-evolution results for the total ($E_K$), compressible ($E^{(c)}_K$), and 
incompressible ($E^{(i)}_K$) kinetic energies 
are presented in Fig.~\ref{fig08}, respectively, for the type-I (left) and 
type-II (right) stirring models. For both cases, the results are displayed in three
panels, considering two frequencies and two DDI parameters, such that 
$(\nu,\alpha)= (0.8\omega_\rho,0^\circ$) [top panels], $(1.5\omega_\rho,0^\circ)$ [middle
panels] and ($1.5\omega_\rho,30^\circ$) [bottom panels]. In these cases, the corresponding
long-time behaviors are kept in the insets. Reminding that the compressible parts
of the kinetic energy, $E^{(c)}_K$, are associated with the sound-wave productions,
with the incompressible ones, $E^{(i)}_K$ related to the vorticity of the fluid and
turbulence, we noticed that in the short-time-interval ($1< \omega_\rho t< 15$) the
compressible part remains increasing slowly, whereas the incompressible part
is increasing much faster, following $E_K$. This can be taken as related to the 
increasing vorticity, with energies being transferred to vortex production and 
turbulence. In this regard, one can also verify that for smaller stirring rotation 
(upper panel), which is related to the vortex-dipole productions, the compressible 
part keeps much lower than the incompressible part even for the longer time 
interval, in contrast with the case that $\nu=1.5\omega_\rho$ (see the middle and bottom 
panels), which is related to the vortex-cluster productions.

By comparing the results shown for the type-I stirring model, in Fig.~\ref{fig08}
(left), with the ones obtained for the type-II model, in \ref{fig08}(right), 
the main difference relies on the increasing amount of kinetic energy, as the
general behavior is similar. Particularly for higher velocities, we notice a 
significant increase in the compressible part of the kinetic energy, which is due 
to the vibration of the Gaussian obstacle.

\subsection{Quantum turbulence: Kolmogorov's energy spectrum}
The emergence of a scaling law in the kinetic energy spectrum is being analyzed 
through our displayed results presented in Fig.~\ref{fig09}, for two
types of stirring models (I and II), where we are
verifying that the classical Kolmogorov behavior $k^{-5/3}$ can be characterized in 
the initial time interval $t\lesssim 15\omega_\rho^{-1}$, with the kinetic energy spectra being
averaged over 10 samples of the vortex regime evolution. As shown, the power-law 
behavior $k^{-5/3}$ is modified to $k^{-3}$ when going to the ultravioled regime,
 which is in agreement with the energy spectra study of vortex 
distributions in 2D quantum turbulence provided in Ref.~\cite{2012Bradley}. 
In view of the similarity with the counterpart classical scaling law behavior, 
we understand these results are quite indicative of quantum turbulence for dipolar
BEC in the dynamics of the vortex-antivortex production. 
The present results provide further support to the characterization of quantum
turbulence in superfluid~\cite{1995Frisch}, which has been found when considering 
simulations of a nonlinear Schrödinger equation in correspondence with the
previously known Navier-Stokes equation solutions for low-temperature 
superfluid and incompressible viscous fluids~\cite{Nore1997}. As we have
essentially shown is that Kolmogorov's scaling behavior (recognized as a
fundamental concept in classical turbulence) is also emerging in a dynamic
spectral analysis of a dipolar BEC under stirring circular interaction, providing
further support to quantum turbulence, as a description of energy distribution 
in turbulent flows. 

\section{Conclusions}
\label{sec5}
By considering a quasi-2D trapped dipolar Bose-Einstein condensate submitted 
to circularly moving Gaussian obstacle (simulating a laser stirring perturbation), we are 
reporting the occurrence of quantum turbulence within the dynamical process of 
vortex-antivortex pair emmission. The pieces of evidence for quantum turbulence are provided 
by the momentum spectral analysis of the incompressible kinetic energy, in which it was verified 
the expected characteristic classical Kolmogorov power-law behavior for turbulence, $k^{-5/3}$, 
in a momentum interval such that $0.2\lesssim k\xi\lesssim 7$ (where $\xi$ is the healing length).
As it happens in the classical fluid dynamics counterpart, the spectral power-law behavior 
changes to $k^{-3}$ when going to the ultraviolet regime. This dynamical process occurs within 
a period of time just after the initial production of vortex-antivortex pairs 
($t\lesssim 15\omega_\rho^{-1}$).
Two variants of stirring dynamics are assumed for the moving Gaussian-shaped penetrable obstacle 
in our model approach (type-I and type-II), being applied to a dipolar condensate confined 
by a quasi-2D pancake-like harmonic trap. For the type-I model, the obstacle moves with constant
rotational frequency at a fixed given radius inside the condensed fluid, with its amplitude 
$A_0$ assumed to be close to 90\% of the stationary chemical potential $\mu$. For type-II model,
we kept the same conditions, except that the amplitude is vibrating with frequency larger than 
the stirring rotational one, in order to verify the effect of an additional dynamics provided 
by the obstacle. Once given the nonlinear (contact and dipolar) interactions, the 
rotation frequency is the main variable to be considered for the dynamics inside the dipolar 
quantum fluid. 

The critical velocities for the nucleation of vortex-antivortex pairs, in both model approaches,
are established by solving the corresponding nonlocal two-dimensional GP equation in real-time.
For higher rotations of the obstacle, a second transition in the dynamics is also verified  
with the production of vortex clusters (identified when more than one pair emerges at each time
within the rotation cycles).
By assuming fixed repulsive contact interactions between the atoms, the critical velocities are
verified with respect to the dipole orientation angle $\alpha$, which can alter the DDI from
positive (maximum at $\alpha=0$) to negative values ($90^\circ\ge \alpha\ge 54.7^\circ$). 
However, restricted by the fixed model parameters, as the contact interaction and radial position 
of the obstacle, we choose in the present simulations $\alpha$ values compatible with repulsive 
dipole-dipole interactions ($\alpha< 54.7^\circ$), such that the dynamical study on 
vortex-antivortex emission remains under the same conditions. 

To illustrate the interplay between contact and DDI in the dynamics, as well as to 
verify the role of DDI in our investigations, we select one case in which the repulsive
DDI is at the maximum ($\alpha=0$, with the previously fixed contact interaction, 
$a_s=50a_0$), for comparison with another case in which the 
DDI is completely removed ($\alpha = 54.7^\circ$, but compensating the missing
repulsive interaction with a larger scattering length, $a_s =177a_0$). 
As our simulations show, 
the vortex and antivortex emerge dynamically in almost identical form in both cases,
slightly faster in case of non-zero DDI, indicating 
the critical velocities are mainly due to the condensate stationary observables as the
chemical potential and rms radius.
However, even before one cycle is completed, a quite different dynamics inside
the fluid is revealed when comparing the two cases, which becomes more obvious 
for longer-time evolution,
reflecting the kind of atom interactions. 
As verified, the DDIs are responsible for more fluctuations 
inside the fluid density, affecting the propagation of the vortex pairs.

The dynamics of vortex-antivortex production are further explored by 
spectral analysis, with the characterization of turbulent dynamics, which
is verified from the initial time interval of the vortex emission regime, when
the incompressible and compressible parts of the kinetic energy start deviating
from each other (near $t\approx 2\omega_\rho^{-1}$ in our simulations).
In this study, we have first considered in detail the long-time evolution 
of the kinetic energy, by separating the corresponding compressible and
incompressible parts. From classical fluid dynamics, it is understood that
vortex tangles are usually signatures of turbulence associated with the 
flow of incompressible viscous fluids. Therefore, by concentrating our
spectral analysis on the incompressible kinetic energy part, obtained by
averaging over several samples in the time evolution, the characterization 
of the turbulence behavior was established by verifying that the
incompressible kinetic energy $E^{(i)}(k)$ follows approximately the classical 
Kolmogorov power-law $k^{-5/3}$~\cite{Kolmogorov} in the momentum region 
$k\xi\lesssim 5$ (where $\xi$ is the healing length), changing to $k^{-3}$ as 
$k$ goes to the ultraviolet region, consistent with previous studies~\cite{2012Bradley}.
Our results are presented by considering stirring rotational frequencies associated
with condensate regions at which we have vortex-antivortex (vortex-dipoles) pairs and 
vortex-cluster productions.
For that, we consider low and high rotational velocities $r_0\nu$ (with $r_0$
fixed) represented by $\nu=0.8\omega_\rho$ and $\nu=1.5\omega_\rho$, respectively, 
using two values for the angle $\alpha$ that provides repulsive DDI strengths:
$\alpha=0^\circ$ (DDI maximized) and $\alpha=30^\circ$. 
As clarified, our simulations have contemplated only repulsive DDI, through
the angle $\alpha$, restricted by the other model parameters, as the contact
atom-atom interaction and the fixed radial position of the obstacle. However,
in principle one can also investigate the dynamics with attractive DDIs,
by shifting the contact interactions to larger values, as indicated by the 
example we are providing in our simulations when the DDI is set to zero.
Together with a more detailed investigations on the critical velocities
under different combinations of the contact and dipolar atom-atom 
interactions, these are possible straight investigations that 
can be done, even before considering beyond mean field effects.

\noindent {\bf Acknowledgement}
L.T. thanks fruitful related discussions with A. Gammal.
We acknowledge partial support from Fundação de Amparo à Pesquisa do Estado de 
São Paulo (FAPESP) [Contracts No. 2020/02185-1 (S.S.) and No. 2017/05660-0 (L.T.)], 
Conselho Nacional de Desenvolvimento Científico e Tecnol\'ogico (CNPq) 
(Procs. 304469-2019-0  and 464898/2014-5) (L.T.), and  
Marsden Fund (Contract No. UOO1726) (R.K.K.). 

\bibliographystyle{amsplain}

 \end{document}